\documentclass[aps,prl,twocolumn,showpacs,superscriptaddress]{revtex4-2}
\usepackage[english]{babel}
\usepackage{graphicx, color}
\usepackage{hyperref}
\definecolor{link}{rgb}{0.1,0.1,0.9}
\hypersetup{colorlinks=true,linkcolor=link,citecolor=link,urlcolor=link,linktocpage}
\usepackage{epstopdf}
\usepackage{csquotes}
\usepackage{color}
\usepackage{amsmath}
\usepackage{amssymb}
\usepackage{longtable}
\usepackage{mathtools}
\usepackage{float}
\usepackage{subfigure}
\usepackage{ulem}
\usepackage{enumitem}
\usepackage{adjustbox}
\usepackage{multirow}

\setcitestyle{numbers,square}

\begin{document}
\title{New Methods for Critical Analysis: Revealing the Simultaneous Existence of Universality Classes in Nontrivial Magnetic Systems}

\author{Harish Chandr Chauhan}
\affiliation
{Department of Physics, Indian Institute of Technology Delhi, New Delhi 110016, India}
\email{Harish.chandr.chauhan@physics.iitd.ac.in}
\email{pintu@physics.iitd.ac.in}
\author{Umesh C. Roy}
\affiliation {Department of Chemistry, College of Staten Island, Staten Island, NY 10314, USA}
\author{Shovan Dan}
\affiliation{Department of Condensed Matter Physics and Materials Science, Tata Institute of Fundamental Research, Homi Bhabha Road, Colaba, Mumbai 400005, India}
\author{A. Thamizhavel}
\affiliation{Department of Condensed Matter Physics and Materials Science, Tata Institute of Fundamental Research, Homi Bhabha Road, Colaba, Mumbai 400005, India}
\author{Pintu Das}
\affiliation
{Department of Physics, Indian Institute of Technology Delhi, New Delhi 110016, India}

\begin{abstract}
	
			In magnetic systems, the microscopic constituents exhibit power law behavior near the paramagnetic transition temperature, $T_{\rm{C}}$. The critical exponents (CEs) associated with the physical quantities that demonstrate singular behavior at $T_{\rm{C}}$ illustrate the critical behavior, specifically the range and type of exchange interactions emerging in magnetic systems. However, it is realized that the developed methodologies may not yield accurate values of CEs, especially for magnetic systems with competing interactions, referred to as nontrivial magnetic systems. Currently, no comprehensive method effectively addresses the competing effects of the range of magnetic interactions among the constituent entities emerging in such systems. Additionally, there is no definitive explanation for CE values that do not belong to any single universality class. Here, we present new methodologies for critical analysis aimed at determining both the range of exchange interaction(s) and appropriate values of CEs. Using computational and experimental investigations, we analyze the magnetic behavior of trivial Ni and nontrivial Gd. Our findings demonstrate that (i) the critical behavior remains the same on either side of $T_{\rm{C}}$, (ii) the critical behavior associated with local electron moments remains unaffected by the magnetic field, and (iii) in Gd, the critical role of competing interactions becomes evident: local electron moments follow a three-dimensional Ising-type short-range interaction, while itinerant electron moments exhibit a mean-field-type long-range Ruderman–Kittel–Kasuya–Yosida (RKKY) interaction, which weakens under an external magnetic field due to the localization effect on itinerant electrons. 
\end{abstract}
\maketitle

\hspace{6 mm} In the studies of phase transitions for cases, such as condensation of gases, melting of solids, magnetic ordering in solids, etc., the understanding of the behavior of the systems adjacent to the critical point is of fundamental interest \cite{RevModPhys.25.191.2, PhysRevLett.98.056601, PhysRevB.100.165143}. Generally, the physics of phase transition is governed by the behavior of the microscopic constituents which interact strongly in a cooperative fashion. Critical exponents (CEs), which define the power law behavior of physical quantities exhibiting singularities at the critical points, indicate the qualitative nature, i.e., the critical behavior of a system \cite{RevModPhys.46.597, RevModPhys.71.S358}. The types of cooperative interactions (short-range, long-range, etc.) among the constituent entities play a significant role in defining the CEs. In magnetic systems, the magnetic moments are known to interact typically via different types of exchange interactions resulting in a set of CEs~\cite{PhysRevB.96.144429, Bisht_2023, J_Geldenhuys_1978, PhysRevB.87.125401, PhysRevB.105.104402, tiwari20203d, PHAN2010238}. Different methodologies have been suggested for the determination of a set of CEs ($\alpha, \beta, \gamma$, $\delta$, etc.) corresponding to physical quantities such as specific heat ($C_{\rm{v}}$), spontaneous magnetization ($M$), susceptibility ($\chi$) and magnetization isotherm recorded at the paramagnetic transition temperature ($T_{\rm{C}}$), respectively for various universality classes (UCs), viz., Heisenberg, XY, Ising, mean field (MF), or tricritical MF~\cite{RevModPhys.46.597, RevModPhys.71.S358}. The CEs are mutually related and they depend critically on the dimensionality (of space and spin) and the range of microscopic interactions.

Several experimental methodologies have been developed to investigate the critical behavior of magnetic systems~\cite{PhysRevLett.19.786, PhysRev.136.A1626, franco2006field}. Although, most methodologies suggest same CEs for two sides of a transition, a recent theoretical work suggested otherwise~\cite{PhysRevLett.115.200601}. In fact, the recent proposal~\cite{PhysRevLett.128.015703} based on a modified Arrott-plot (MAP)~\cite{PhysRevLett.19.786} analysis, (i) pointed out that the existing methodologies for critical analysis may yield ambiguous values of CEs, (ii) proposed a new method for critical analysis, and (iii) supported the claim of ref.~\cite{PhysRevLett.115.200601}. Although many important points have been settled for the study of critical behavior, we are still left with several crucial open problems, as discussed below, which are needed to be resolved to fully understand the critical behavior of magnetic systems.

The critical behavior of numerous nontrivial magnetic systems, such as ${\mathrm{La}}_{0.6}{\mathrm{Ag}}_{0.2}{\mathrm{Bi}}_{0.2}{\mathrm{MnO}}_{3}$ \cite{Bisht_2023}, ${\mathrm{La}}_{0.7}{\mathrm{Sr}}_{0.3}{\mathrm{MnO}}_{3}$ \cite{PhysRevLett.81.4740}, ${\mathrm{Co}}_{3.6}{\mathrm{Fe}}_{4.4}{\mathrm{Zn}}_{8}{\mathrm{Mn}}_{4}$ \cite{BHATTACHARYA2023170274}, ${\mathrm{La}}_{0.75}{\mathrm{Sr}}_{0.25}{\mathrm{MnO}}_{3}$ \cite{PhysRevB.65.214424}, and ${\mathrm{Gd}}$, etc. \cite{PhysRevB.60.12166, PhysRevB.33.4747, PhysRevLett.62.2728, 10.1063/1.5093555} have been reported in literature. Here, the magnetic systems with competing interactions are referred to as nontrivial. In most of the cases, it was highlighted that the reported CE values in nontrivial magnetic systems do not belong to any single UC \cite{Bisht_2023, PhysRevLett.81.4740, BHATTACHARYA2023170274, PhysRevB.65.214424, PhysRevB.60.12166} which is a serious impediment in the understanding of critical behavior of magnetic systems in general. The answer to this observation remains elusive; therefore, a deeper understanding of critical behavior in nontrivial magnetic systems is crucial and requires more thorough investigation. Moreover, the value of $\gamma$ obtained using the developed methodologies is essentially used to determine $\sigma$ \cite{PhysRevLett.29.917} which defines the range of interaction in magnetic systems. The theoretical suggestions for long- and short-range interactions correspond to $\sigma\leq 1.5$ and $\sigma\geq2.0$, respectively, which often leads to conflicting interpretations, particularly when $\sigma$ falls between the two values \cite{RevModPhys.46.597, RevModPhys.71.S358, PhysRevB.8.281, PhysRevB.15.4344, tiwari2021critical, BHATTACHARYA2023170274}. We note here that the $\gamma$ estimated for nontrivial magnetic systems is an effective value. Therefore, an inaccurate determination of $\sigma$ $-$ which effectively depends on $\gamma$ \cite{PhysRevLett.29.917} $-$ is likely to lead to nonphysical conclusions. This highlights the lack of a reliable methodology for accurately identifying the range of competing interactions, especially in nontrivial systems. The question of whether the two sets of exponents on either side of $T_{\rm{C}}$ are due to the intrinsic complex interactions or due to technical issues in the methodologies used for the analysis of magnetic behavior requires a thorough investigation. Furthermore, the influence of external magnetic field ($B=\mu_0H$) on the competing interactions remains an open problem to investigate. On the technical front, a new methodology is a prerequisite to determine the accurate values of CEs in least number of iterations.

In this letter, we present our detailed computational and experimental investigations of the critical behavior that potentially emerge in trivial as well as nontrivial magnetic systems. For computational investigations, we chose Ni as a trivial system. (i) We report a novel methodology to determine precise and accurate values of the CEs in least number of iterations. Using the proposed method, robust and convergent results are obtained for Ni. (ii) We propose a new methodology to qualitatively identify the range of exchange interaction in these systems. After confirming the correctness of the results for Ni using our developed methodologies, we investigate the critical behavior of nontrivial single crystalline Gd. (iii) With our new approach of analysis of experimentally determined magnetization isotherms, we identify that the local and itinerant electron moments in Gd exhibit different types of exchange interactions following different UCs thereby resulting in effective values of CEs. (iv) We also analyze the effect of $B$ on the nature of the competing interactions in nontrivial systems. (v) Subsequently, we show that magnetic systems hold identical values of the CEs on either side of $T_{\rm{C}}$.

\begin{figure}
\centering{}
\includegraphics[width=\linewidth]{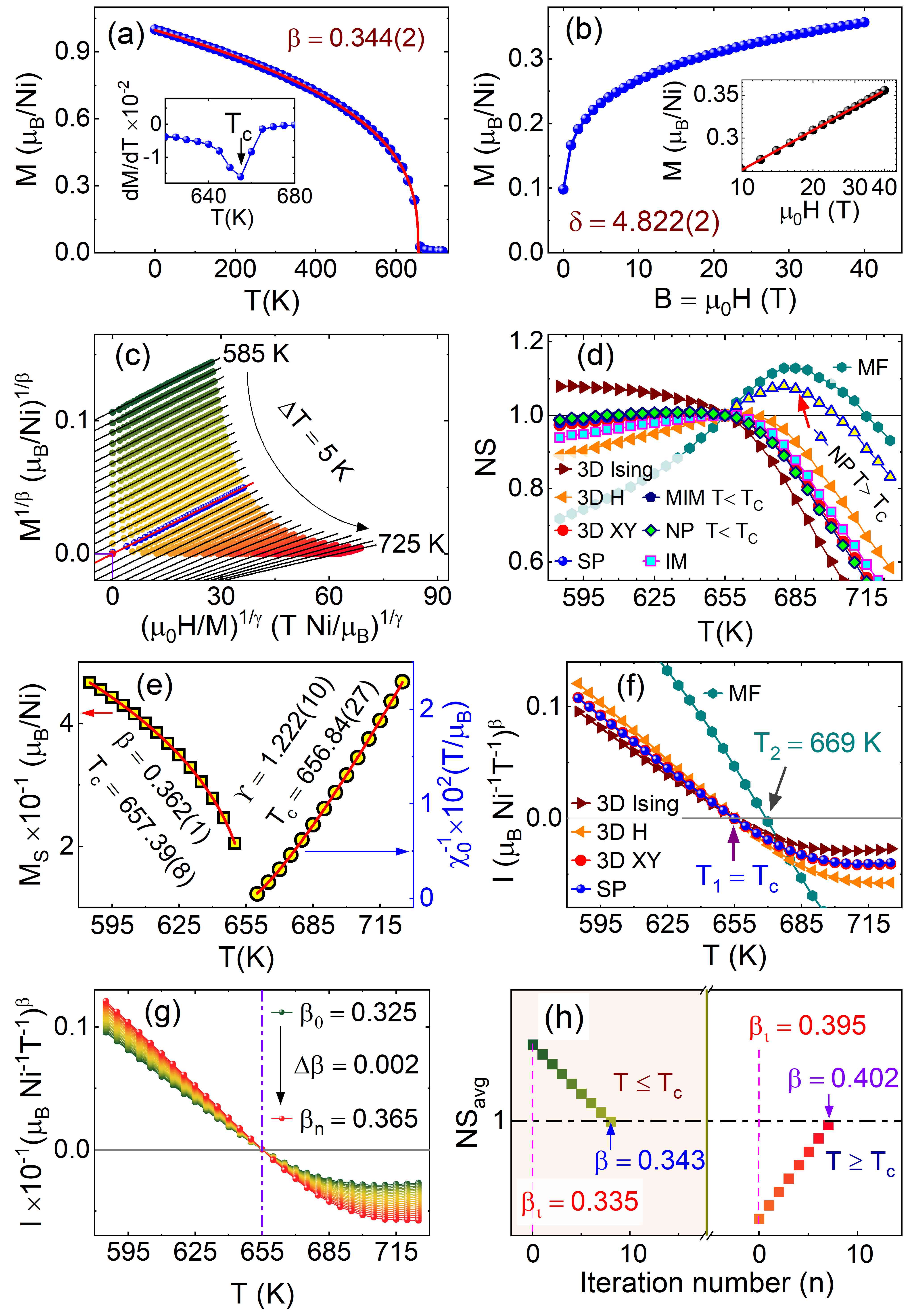}
\caption{(a) The $M-T$ plot obtained using the Vampire atomistic simulation for Ni. Inset of (a) represents the first derivative of the $M-T$ resulting $T_{\rm{C}}=655$ K. (b) The $M-H$ isotherm estimated at $T=T_{\rm{C}}$. Inset shows the linear fit to log-log plot of $M-H$. (c) The MAPs constructed with CEs determined using SP. (d) $NS$ vs. $T$ plots constructed using the CEs of the standard UCs, and the CEs determined from IM, MIM, SP and the new proposal (NP). (e) The plots of $M_S$ and $\chi_0^{-1}$ against $T$. Symbols represent simulated data while solid lines are fit to respective curves. (f) $I$s vs. $T$ plots constructed using the CEs determined using SP and of standard UCs. (g) $I$s vs. $T$ plots for different values of $\beta$ and $\gamma$ which are related as $\gamma = \beta\times(\delta-1)$. (h) Variation of $NS_{\rm{avg}}$ with $n$ which is related with $\beta$ as $\beta(n)=\beta_i+\Delta\beta\times n$. Inset of (h) represent the $NS_{\rm{avg}}$ vs. $\beta$ plots constructed using a few sets of $\beta$ and corresponding $\gamma$ as mentioned in (g). MF: mean field, TMF: tricritical MF and H: Heisenberg.}
\label{MT_MH_IM_Ni}
\end{figure} 

In order to determine precise values of CEs, we first proceed with the analysis of the magnetization ($M$) data for Ni. The temperature ($T$)-dependent magnetization ($M-T$) for both zero- and applied-$B$ are determined by carrying out atomistic simulations based on solution of Landau-Lifshitz-Gilbert equation using Vampire simulation package~\cite{Evans_2014, SM}. $T_{\rm{C}}=655$\,K is determined from the minima of the first-derivative [inset of Fig. \ref{MT_MH_IM_Ni}(a)]. Using the relations, $M\propto|\epsilon|^{\beta}$, $M\propto(\mu_0H)^{\frac{1}{\delta}}$ at $T=T_{\rm{C}}$ and the Widom relation $\delta=1+\frac{\gamma}{\beta}$ ~\cite{RevModPhys.71.S358, RevModPhys.46.597, ME_Fisher_1967, widom1974critical}, we obtain $\beta=0.344(2)$ [Fig. \ref{MT_MH_IM_Ni}(a)], $\delta=4.822(2)$ [Fig. \ref{MT_MH_IM_Ni}(b)] and $\gamma=1.315(8)$, respectively, which corresponds to 3D-XY UC. Here, $\epsilon=\frac{T-T_{\rm{C}}}{T_{\rm{C}}}$ is the reduced $T$. The above methods of determination of CEs are identified as the standard procedure (SP). We note here that in many real systems, SP may not be always applicable for critical analysis, see discussions below. In the following, we analyze the primary issues involved with the other existing methods for critical analysis such as IM, modified IM (MIM)~\cite{PhysRevLett.128.015703}, etc. to determine the critical behavior of magnetic systems.

A. Arrot and J. E. Noakes \cite{PhysRevLett.19.786} suggested a linear relationship between $M^{1/{\beta}}$ and $(B/M)^{1/{\gamma}}$ for appropriate values of CEs. The plots of $M^{1/{\beta}}$ vs. $(B/M)^{1/{\gamma}}$ is termed as MAPs, see Fig.\,\ref{MT_MH_IM_Ni}(c). The linearity of the MAPs plotted using the values of $\beta$ and $\gamma$ can be checked by determining the normalized slopes ($NS$s), defined as slopes of the MAP at a given $T$ normalized with respect to that at $T=T_{\rm{C}}$. Figure\,\ref{MT_MH_IM_Ni}(d) shows that the values of $NS$, if determined using the correct CE values obtained by the SP, are $\sim 1$ for $T\leq T_{\rm{C}}$. We, therefore, use $NS\sim1$ for $T\leq T_{\rm{C}}$ as the criterion to test the validity of a method of critical analysis.  We note here that for $T>T_{\rm{C}}$, the values of $NS$ deviate significantly from $1$ which we relate to choosing the appropriate range of $T$ for the analysis of critical behavior of a system ~\cite{PhysRevLett.128.015703, SM}, see analysis and discussion on this below. 

In order to check the validity of IM for critical analysis, we determine the CE values for Ni using the IM. The values of CEs, viz. $\beta = 0.362(1)$ and $\gamma = 1.222(10)$ [Fig.\,\ref{MT_MH_IM_Ni}(e)] obtained by IM are markedly different from the ones obtained using SP. Moreover, as shown in Fig.\,\ref{MT_MH_IM_Ni}(d), the $NS$s for MAPs plotted using these values of $\beta$ and $\gamma$ deviate significantly from the criterion $NS\sim 1$ on either side of $T_{\rm{C}}$ demonstrating that the IM may not be an appropriate approach for critical analysis of a magnetic system. Interestingly, the CEs ($\beta=0.343(3)$, $\gamma=1.309(8)$) determined using MIM and $NS\sim1$ for $T\leq T_{\rm{C}}$ shows 3D XY type short-range ($\sigma\approx2$) interaction~\cite{BHATTACHARYA2023170274, SM}. However, appropriate values of CEs for $T\geq T_{\rm{C}}$ could not be determined using MIM as $NS$ values deviate significantly from $1$, see \cite{SM}. Thus, it is clear that the existing methodologies may not provide robust results which has a profound effect on the general understanding of phase transitions of a system. Below, we propose novel methodologies for the analysis of critical behavior of magnetic systems and demonstrate with examples of Ni and Gd that the criterion we discussed above is fully satisfied.

Figure \ref{MT_MH_IM_Ni}(f) shows the plots of the $T$-dependence of \textit{intercepts} ($I$s) obtained from the linear fits of MAPs for Ni corresponding to different UCs. It is clear that the $I$s, estimated by both SP as well as by using the CEs of short-range standard UCs, show a crossover at $T\sim T_{\rm{C}}$. However, if $I$s are estimated considering MF UC representing a long-range interaction, the crossover $T$ ($T_2=669$\,K) deviates from $T_{\rm{C}}$. Since the short-range magnetic interaction is established for Ni from the SP of critical analysis, we identify the crossover of the $I$s at $T=T_{\rm{C}}$ as the criterion to determine the range of interaction in a magnetic system. The same approach may also be applicable to identify long-range interactions if the $I$s, determined using the CEs of MF UC, show a crossover around $T_{\rm{C}}$. We show below that this new method of analysis based on $I$s of MAPs may be a proper approach to determine the appropriate range of interactions. By using this approach, we have demonstrated below the role of itinerant electrons in the magnetic behavior of Gd. Moreover, we note that the different sets of $\beta$ and $\gamma$ values corresponding to different short-range standard UCs lead to the same value of $\delta$ as determined from the $M-H$ at $T_{\rm{C}}$, see Fig. \ref{MT_MH_IM_Ni}(b). Additionally, the values of the $I$s of MAPs determined using the CEs corresponding to 3D Ising, 3D XY, 3D Heisenberg and SP converge to $0$ at $T=T_{\rm{C}}$ (the crossover $T$). Thus, although the type of interaction (i.e., short-range) has been identified, the exact nature of short-range interaction is still to be determined.

To determine the accurate values of the CEs, we proceed as follows. We vary $\beta$ in a controlled way such that $\beta(n)=\beta_i\pm \Delta\beta\times n$, where $\beta_i$ is the initially selected value, $\Delta\beta$ is the increment in $\beta$ for each iteration and $n$ is the iteration number. To start with, $\beta_i$ can be chosen as a theoretical value corresponding to any standard UC or even any random value with the condition that the chosen set of $\beta$ and corresponding $\gamma$ lead to linear behavior of MAPs, see Fig. \ref{MT_MH_IM_Ni}(c) for example. The corresponding values of $\gamma$ are determined for each $\beta(n)$ using the Widom relation~\cite{widom1974critical}, $\gamma=\beta(n)\times(\delta-1)$, where $\delta$ is directly obtained from $M-H$ curve at $T=T_{\rm{C}}$ [Fig.\, \ref{MT_MH_IM_Ni}(b)]. The $I$s of MAPs thus obtained are found to exhibit a crossover at $T_{\rm{C}}$ [Fig.\,\ref{MT_MH_IM_Ni}(g)] for multiple sets of $\beta$ and $\gamma$ values. In order to determine appropriate values of $\beta$ and $\gamma$, the MAPs obtained as above are used to estimate an average value ($NS_{\rm{avg}}$) of $NSs$ for $T\leq T_{\rm{C}}$ and $T\geq T_{\rm{C}}$ separately. $NS_{\rm{avg}}$ is defined by $NS_{\rm{avg}}=\frac{\sum_{i=1}^{N} (NS)_i}{N}$, where $i$ denotes the $M-H$ isotherm at a given $T$ and $N$ is the total number of the isotherms used in the analysis. The $NS_{\rm{avg}}$ thus determined for each iteration $n$ are plotted in Fig. \ref{MT_MH_IM_Ni}(h). According to our proposal, the set of CE values ($\beta$, $\delta$ and $\gamma$) corresponding to $NS_{\rm{avg}}=1$ and $NS\sim 1$ are the appropriate values of the CEs for a system. Using our method proposed here, we determine $\beta=0.343(3)$ and $\gamma=1.311(18)$ for $T\leq T_{\rm{C}}$ which are remarkably consistent with the CEs obtained using the SP. The error in $\beta$ is manually controllable while the error in $\gamma$ depends on the imposed error in $\beta$ and the error appearing in $\delta$.

However, for $T\geq T_{\rm{C}}$, we find that the values of CEs ($\beta=0.402(5)$ and $\gamma=1.533(20)$), resulting $NS_{\rm{avg}}\sim1$, are significantly different than as obtained for $T\leq T_{\rm{C}}$, see Fig.\,\ref{MT_MH_IM_Ni}(h). Moreover, although $NS_{\rm{avg}}\sim1$ for $\beta=0.402(5)$, the NS values show significant deviation from the criterion of $NS\sim 1$ [Fig. \ref{MT_MH_IM_Ni}(d)]. This suggests that the CEs obtained for $T\geq T_{\rm{C}}$ may not define the critical behavior. This is further supported from $\beta$ which is undefined for above $T_{\rm{C}}$ \cite{ME_Fisher_1967}. Moreover, it is difficult to identify the $B$-dependent $T$-range up to which the short-range exchange interaction survives since the thermal fluctuation mainly control the physics above $T_{\rm{C}}$ in the vicinity of $T_{\rm{C}}$. The later statement indicate that one may get different values of $\gamma$ for below and above $T_{\rm{C}}$ depending on the properties of the chosen system for investigation as suggested in ref. \cite{PhysRevLett.115.200601}. This is because the correlation length $\xi$ is usually different for below and above $T_{\rm{C}}$ in nontrivial magnetic systems. From our detailed analysis, we infer that the consideration of \textit{identical range of} $T$ for the analysis below and above $T_{\rm{C}}$ may have resulted in such a discrepancy. In the following, we verify this and discuss a procedure to estimate the appropriate range of $T$ and $B$ to be used for critical analysis for below and above $T_{\rm{C}}$.

\begin{figure}
\centering{}
\includegraphics[width=8cm]{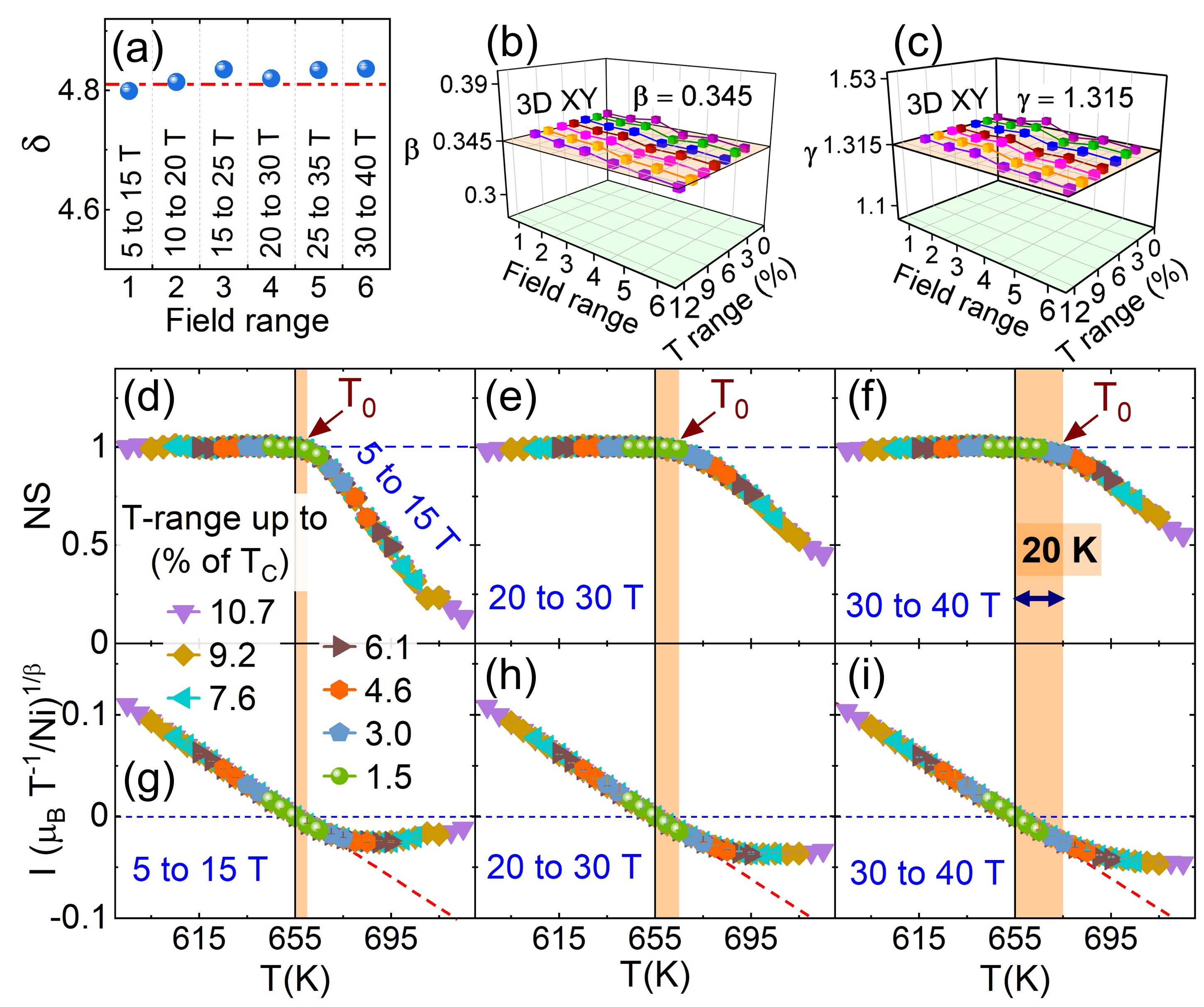}
\caption{(a) The field range dependence of $\delta$. Different field ranges are mentioned in the Figure. (b) and (c) represent the variation of $\beta$ and $\gamma$, respectively, for different $B$ and $T$ ranges determined using our new proposal (NP). (d)$-$(f) and (g)$-$(i) represent $NS$ and $I$ vs. $T$ plots, respectively, constructed using the CEs obtained using NP for different $B$ and $T$ ranges.}
\label{Temperature_and_Field_range}
\end{figure}

Figures \ref{Temperature_and_Field_range}(a, b, c) show the analysis of the behavior of $\delta$, $\beta$ and $\gamma$ in different ranges of $T$ and $B$ as determined using our proposed method, as described above, for $T\leq T_{\rm{C}}$. In Figs.\,\ref{Temperature_and_Field_range}(b-i), seven different $T$ ranges, viz., $10.7$\% of $T_{\rm{C}}$ to $1.5\%$ of $T_{\rm{C}}$ and three different ranges of $B$ are considered for all the relevant parameters discussed above. From Figs.\,\ref{Temperature_and_Field_range}(a-c) we find that the CEs thus obtained are robust across different $T$ (for $T\leq T_{\rm{C}}$) and $B$ ranges clearly suggesting  that the critical behavior associated with local electron moments is independent of $B$. Figures \ref{Temperature_and_Field_range} (d-f) show the representative plots of $NS$s, calculated from MAPs using these robust values of CEs obtained using the proposed method for $T\leq T_{\rm{C}}$. For brevity, three different $B$ ranges, viz., 5-15\,T, 20-30\,T and 30-40\,T, respectively, are shown. Clearly, the $NS\sim 1$ for the entire $T$ range for $T\leq T_{\rm{C}}$. However, for $T>T_{\rm{C}}$, we observe that the values deviate from  $NS\sim 1$ beyond $T=T_0$ [Fig. \ref{Temperature_and_Field_range}(d)] which is very close to $T_{\rm{C}}$. Interestingly for $T>T_{\rm{C}}$, the $T$-range, $T_0-T_{\rm{C}}$ showing $NS\sim1$ grows with increasing $B$ [Figs. \ref{Temperature_and_Field_range}(e and f)]. We thus confirm that the CE values will remain the same on either side of $T_{\rm{C}}$ if one select the temperature range up to $T_0$ for $T>T_{\rm{C}}$. We, therefore, propose a process to determine the appropriate ranges of $T$ and $B$ for critical analysis. Selecting the $M-H$ isotherms at relatively high-$B$ and low-$T$ ranges, $NS$s are to be determined using the $\beta$ and corresponding $\gamma$ as we proposed above. The maximum $T$-range for the chosen $B$ range showing $NS\sim 1$ will be the appropriate range to be considered for the critical analysis. It is to be noted that the $B$-dependent critical analysis is required if $\delta$ varies with $B$. Furthermore, as we verify the behavior of linearity of $NS$ with $T$ [Figs.\ref{Temperature_and_Field_range}(g-i)], we find that the range of linearity of the $I$s also increases with increasing $B$-range. We thus show that the appropriate $T$-range can also be determined from the plot of $I$s vs. $T$ exhibiting the linear behavior.

\begin{figure}
\centering{}
\includegraphics[width=8cm]{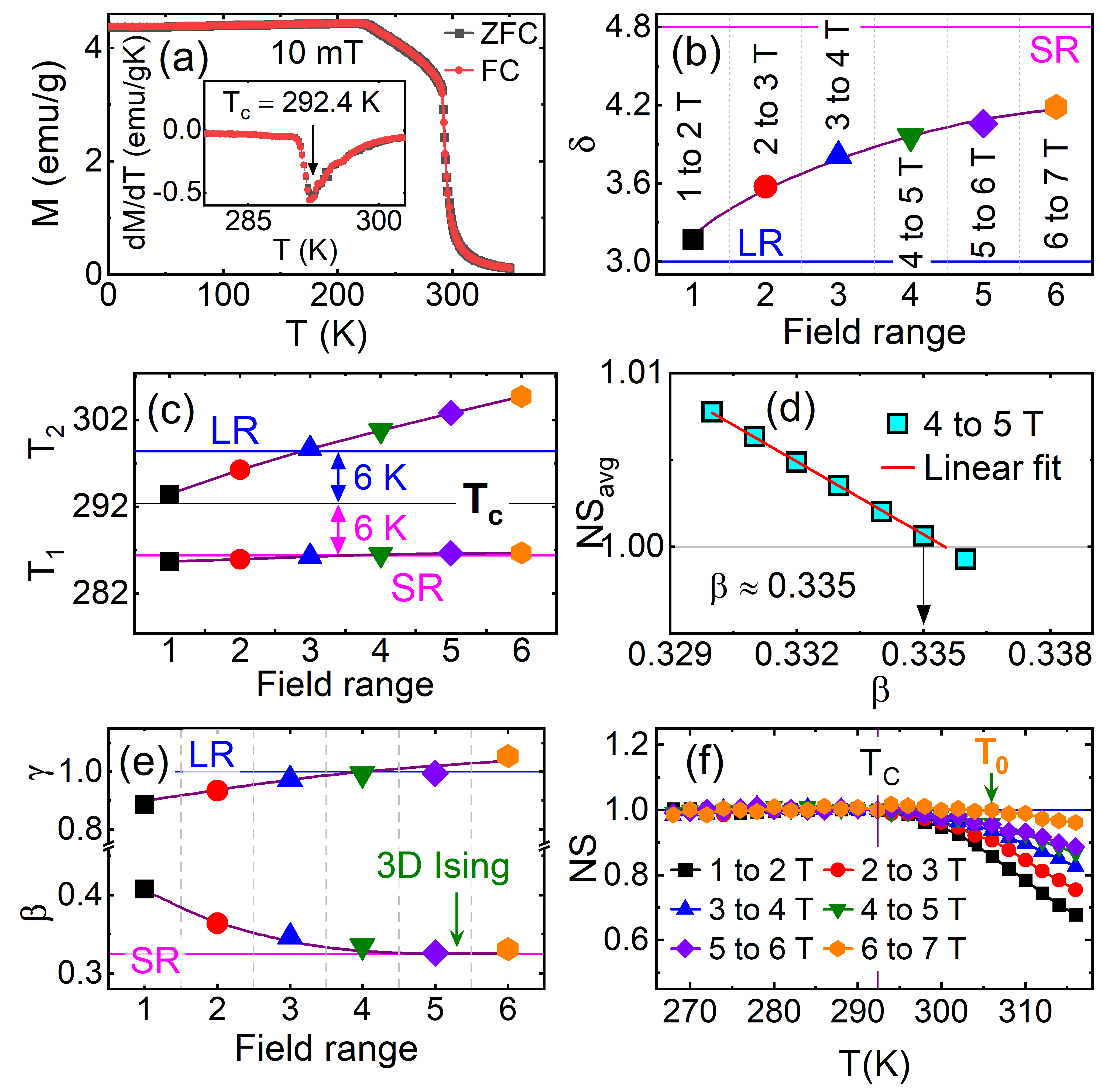}
\caption{(a) The $M-T$ recorded for Gd in field-cooled (FC) and zero-field-cooled (ZFC) warming modes by applying $B=10$ mT along [0001]. Inset is the first-derivative of $M-T$ which yields $T_{\rm{C}}=292.4$ K. (b) The $B$-dependent variation of $\delta_{MH}$, estimated for different $B$ ranges as mentioned. (c) Variation of the crossover temperatures $T_1$ and $T_2$ obtained from the $I$s of the linear fit to MAPs constructed using the CE values of the short-range standard UCs and MF UC, respectively. (d) Plots of $NS_{\rm{avg}}$ vs. $\beta$ to estimate appropriate $\beta$ and $\gamma$ associated with Gd in the $B$-range from 4 to 5 T. (e) Variation of $\beta$ and $\gamma$, determined using our proposal, for different $B$ ranges. (f) $NS$ vs. $T$ plots, constructed using the CEs obtained using our proposal, for different $B$ ranges. SR: short-range, LR: long-range.}
\label{GD_CB}
\end{figure}

We next apply our proposed methods to investigate the role of itinerant electron magnetism on the critical behavior of Gd. Although the magnetic properties of this rare earth material is interesting for applications, however, the exact nature of the  interactions leading to the observed magnetic behavior of Gd is complex and still remains unclear~\cite{PhysRevB.105.104402, PhysRevLett.115.096402, Frietsch2015, PhysRevLett.115.096402, L.M.Sandratskii_1993, KHMELEVSKYI2005145}. The complex competing interactions between local $4f$ as well as itinerant electrons makes Gd an interesting \enquote{nontrivial} magnetic system~\cite{Frietsch2015, PhysRevLett.115.096402, L.M.Sandratskii_1993, KHMELEVSKYI2005145}. The previous studies for critical behavior of Gd report conflicting conclusions as regards to the nature of the magnetic interaction~\cite{PhysRevB.33.4747, PhysRevLett.62.2728, PhysRevB.60.12166, 10.1063/1.5093555}. Although the role of long-range Ruderman-Kittel-Kasuya-Yusida (RKKY) interaction involving itinerant electrons was suggested for Gd~\cite{PhysRevB.105.104402, J_Geldenhuys_1978}, so far the analysis of critical behavior using existing methods indicated the presence of short-range interactions only~\cite{PhysRevB.33.4747, PhysRevLett.62.2728, PhysRevB.60.12166, 10.1063/1.5093555}. From $B$-dependent studies, it was also suggested that the 3D Heisenberg UC in low-$B$ changes to 3D Ising in the high-$B$~\cite{10.1063/1.5093555}. However, the values of $\gamma$ and $\delta$ reported for low-$B$ range do not fully justify the claim. To test our proposed method of critical analysis for this nontrivial magnetic system, we use high-quality single crystalline Gd in hexagonal close packed form, grown by Czochralski method~\cite{TOMASZEWSKI20021, SM}. The magnetization ($M-T$ and $M-H$) measurements were carried out in a SQUID setup (make: Quantum Design) by applying $B$ along [0001]~\cite{SM}. The $T_{\rm{C}}$ for the crystal is 292.4\,K [Fig. \ref{GD_CB}(a)], and the magnetic moment per Gd atom is found to be $\approx 7.63\,\mu_B$ which is consistent with the values reported in literature~\cite{PhysRevB.57.3478}. The unusual characteristic features observed near $T_{C}$ in $M-T$ plot [see Fig.\,\ref{GD_CB}(a)] is ascribed to the competing energies in the system
\cite{Frietsch2015, PhysRevLett.115.096402, L.M.Sandratskii_1993, KHMELEVSKYI2005145, PhysRevB.105.104402}. Due to this, determination of $\beta$ using SP [Figs.\,\ref{MT_MH_IM_Ni}(a) and \ref{MT_MH_IM_Ni}(b)] is not straight forward. This serves as a prerequisite for applying our proposed methods to analyze the critical behavior of Gd.

Unlike for Ni [Fig.\,\ref{Temperature_and_Field_range}(a)], our results for Gd clearly show a variation of $\delta$ with $B$, see Fig.\,\ref{GD_CB}(b). The range of $\delta$ observed in this case covers nearly the values for long-range MF to that for short-range standard UCs. So, we proceed for $B$-dependent critical analysis by considering 8\% of $T_{\rm{C}}$ as the appropriate $T$-range for $T \leq T_{\rm{C}}$. The $I$s of the linear fit to MAPs, constructed using the CEs for short-range standard UCs as well as long-range MF UC, when plotted against the corresponding temperatures exhibit a crossover at $T_1$ and $T_2$, respectively~\cite{SM}. For reference, see Fig.~\ref{MT_MH_IM_Ni}(f) for the corresponding plots for Ni. According to our proposal discussed above, the $I$s of MAPs obtained using CEs of either short-range UCs  or long-range MF UC exhibiting a crossover at $T_{\rm{C}}$ would determine whether the system has a short-range or a long-range interaction, respectively. Our analysis, which shows that $T_2\sim T_{\rm{C}}$ at low-$B$ but deviates from $T_{\rm{C}}$ at high-$B$ [Fig. \ref{GD_CB}(c)], suggests that the long-range interaction mediated by itinerant electron moments weakens with increasing $B$. The Lorentz force acts to localize itinerant electrons, thereby suppressing their long-range interaction characteristics~\cite{PhysRevB.86.041109, KIM1980373}. On the other hand, $T_1$ starting at about $6$\,K away from $T_{\rm{C}}$ [Fig. \ref{GD_CB}(c)] in the low-$B$ range remains almost constant in the entire measured range of $B$ suggesting no effect of $B$ on the range of exchange interaction associated with local electron moments. This competing effect indicates that at high $B$, the interactions are predominantly short-range in nature due to the weakening of the RKKY interaction.

Next, we determine CE values associated with Gd using our new approach, see Fig. \ref{GD_CB}(d) \cite{SM}. Here, $B$-dependent critical analysis is prerequisite as value of $\delta$ changes with $B$. The CE values determined for Gd show results [Fig. \ref{GD_CB}(e)] which are consistent with that obtained from the $I$s [Fig. \ref{GD_CB}(c)] analyses. The $B$-dependent \textit{effective} values of $\beta$ and $\delta$ indicate the dominance of MF type long-range interaction in low-$B$ and predominantly 3D Ising type short-range interaction at high $B$, see Figs.\,\ref{GD_CB}(b) and \,\ref{GD_CB}(e). Moreover, $\gamma$ increases monotonically and found to remain close to 1 in the entire measured range of $B$ [Fig.\,\ref{GD_CB}(e)] indicating the role of a complex competing interactions in Gd. One should note here that the critical analysis yields the effective values of CEs. Interestingly, this has gone largely unnoticed. It remains unclear why the range and type (Heisenberg, Ising, XY, etc.) of exchange interactions would vary with $B$ in a simple ferromagnet. While conventional analyses \cite{PhysRevLett.29.917} indicate predominantly long-range interactions, supported by $\sigma \approx 1.5$~\cite{SM} across the entire field range, our intercept analysis reveals a clear competing effect of short- and long-range interactions with increasing $B$ in Gd. Based on our investigations using the proposed new approaches, we argue that the local Gd moments interact via 3D Ising-type short-range interaction, while the presence of itinerant electrons gives rise to MF-type long-range RKKY interaction. The competition gives rise to effective critical exponent (CE) values that vary with $B$, reflecting the coexistence of multiple UCs arising from distinct exchange interactions. With increasing $B$, the itinerant electrons begin to localize~\cite{PhysRevB.86.041109}, resulting in the suppression of long-range interactions, while the local electron moments remain unaffected, leading to a reduction in the $\beta$ value. Our investigations in this work thus clarifies the reasons behind the CE values, for nontrivial systems, not belonging to any single UC \cite{tiwari20203d, PhysRevB.96.144429, Bisht_2023, PHAN2010238}. Thus, it can be argued that the critical behavior of a magnetic system heavily depends on the local environment around the magnetic ion(s) \cite{PhysRevLett.81.4740, PhysRevB.65.214424, PhysRevB.96.144429, PHAN2010238}. If the local environment of the magnetic ions changes with the field, the CE values will also vary with the field, and vice versa. We again check the robustness and reliability of the obtained CE values (for $T\leq T_{\rm{C}}$) for Gd from the $NS$ plots which show $NS\sim 1$ in the entire $T\leq T_{\rm{C}}$ and in a small range of $T$ above $T_{\rm{C}}$, see Fig. \ref{GD_CB}(f). Similar to the case of Ni, the range of $T$, showing $NS\sim1$ beyond $T_{\rm{C}}$, grows with $B$ indicating further that the critical behavior remains same on either side of $T_{\rm{C}}$.

In conclusion, we propose new methods to determine two important parameters for the analysis of critical behavior of magnetic systems, viz., the range of magnetic interactions and the values of CEs. We demonstrate the successful application of the methods for two magnetic systems, viz. Ni and Gd. By analyzing the $I$s of MAPs, we appropriately identify the range of exchange interaction in a system. From computational and experimental investigations of the critical behavior of Ni and Gd, we claim that $NSs\sim1$ for all $M-H$ isotherms for  $T\leq T_{\rm{C}}$ is the necessary and sufficient condition to test the reliability and robustness of the CE values as determined using our proposed method. We argue that the magnetic systems with competing interactions may simultaneously exhibit two types of UCs resulting in effective values of CEs. Our investigations show that the strength of short-range interaction among local moments remains same while the RKKY type long-range interaction emerging in Gd, due to the involvement of itinerant electron moments, weakens with increasing $B$. These investigations thus solve the long-standing problem of the values of CEs not belonging to any specific UC. The methods and approaches discussed in this article is equally applicable for all types of magnetic systems holding the necessary conditions required for critical analysis. Our proposed methods may thus enable a detailed understanding of the critical behavior of magnetic systems.

\section*{ACKNOWLADGEMENTS}
\hspace{6 mm} We acknowledge the Department of Physics, Indian institute of technology Delhi, New Delhi for providing the MPMS facility for magnetic measurements.  This work is supported by Indo-French Centre for the Promotion of Advanced Research (IFCPAR/CEFIPRA) through grant no. 6508-2/2021 and Science and Engineering Research Board, Govt. of India through grant no. SPR/2021/000762.

\newpage
\renewcommand{\thefigure}{S\arabic{figure}}
\setcounter{figure}{0} 

\paragraph{\large\bf{Supplemental Materials}}\

This Supplemental materials includes\\
1. Materials and Methods\\
2. Modified Arrott-plots\\
3. Iteration method\\
4. Modified Iteration method\\
5. New proposals\\
6. Normalized slope plots for Ni\\
7. Field-dependent normalized slope and intercept plots for SC Gd\\
8. Modified Arrott-plots: theoretical and experimental results\\
9. Scaling results\\
10. Table of critical exponents (CEs) of standard 3D models, and the CEs determined for Ni and SC Gd using various methods

\paragraph{\large\bf{1. Materials and Methods}}\

\textbf{a. Computational}

Face centered cubic (FCC) Ni system was taken for the temperature- and field-dependent magnetization calculation. All the calculations were performed with the vampire software \cite{PhysRevB.91.144425, Evans_2014} using the atomistic model simulations of magnetic Ni. The atomistic Landau–Lifshitz–Gilbert (LLG) equation was solved using the Heun integration scheme to calculate the magnetization curves \cite{PhysRevB.91.144425}. The details about the LLG-Heun method can be found in \cite{Evans_2014, PhysRevB.91.144425}. For the calculations for FCC Ni system, we take the following parameters: cluster size of 20$\times$20$\times$20 $nm$, unit cell size of 3.524 \AA, atomistic spin moment $\mu_S=0.606$$\mu_B$, exchange energy $J_{ij}=2.757\times 10^{-26}$ J/link and we used magnetocrystalline anisotropy energy $k_u=0.0$ J/atom. For the LLG simulation, we use 50000 equilibrium steps and 500000 averaging steps. For all the calculations, we use the critical value of the Gilbert damping parameter equal to $1.0$ and the computational time step equal to $0.1$ fs \cite{Evans_2014, PhysRevB.91.144425, cryst13030537}. We calculated temperature-dependent magnetization (M-T) with and without applying magnetic field. Finally, we converted the M-T isofields into field-dependent magnetization (M-H) isotherms and, thus, performed the theoretical investigation of the critical behavior of Ni system using the methods described below in sections $\mathrm{III}$ to $\mathrm{V}$.

\textbf{b. Experimental}

A single crystal (SC) of gadolinium (Gd) was grown using the Czochralski
method \cite{TOMASZEWSKI20021} in a tetra-arc furnace (Technosearch Corporation, Tokyo, Japan).
Approximately 7 grams of 99.9\% pure Gd metal was placed on a
water-cooled copper hearth inside the furnace. After evacuating the
chamber to a vacuum of 10$^{-6}$ mBar, it was filled with high-purity
argon gas. The Gd metal was melted using the conventional arc melting
process. After the initial melt, the Gd was flipped and remelted under
the same argon atmosphere. Gd melts at around 1300 °C. A
previously prepared polycrystalline Gd seed crystal was carefully
introduced into the molten Gd, causing local solidification around the
seed. The seed crystal was then rapidly pulled at a speed of
approximately 50 mm/h. The melt temperature was gradually increased to
initiate the necking process, after which it was gradually decreased to
widen the diameter of the growing crystal. Once steady-state conditions
were achieved, the crystal was pulled at a steady speed of 10 mm/h for
around 7 hours. The pulled ingot was subsequently aligned along its
principal crystallographic directions using Laue diffraction and cut,
using a wire EDM, along the basal plane and $c$-axis for anisotropic
studies. Since the crystal was prone to oxidation in ambient conditions,
precautions were taken to minimize its exposure to the atmosphere.

\hspace{6 mm} Temperature- and field-dependent magnetic (M-T and M-H) measurements of SC Gd were carried out using Quantum Design magnetic properties measurement system (MPMS). The magnetic field was applied along [0001] direction.

\textit{Temperature-dependent magnetization (M-T)scan in zero-field cooled warming (ZFCW) mode}: First, the sample was cooled to desired low-temperature, the M-T data were recorded with increasing temperature.

\textit{M-T scan in field cooled warming (FCW) mode}: The sample was taken to high enough temperature in the paramagnetic region without applying field. One can guess the paramagnetic region from the recorded M-T in ZFCW mode. Further, after applying field, sample was cooled to desired low-temperature. The M-T data was recorded in warming mode.

\textit{Field-dependent magnetization (M-H) scan}: Without applying field, the sample was cooled to desired low-temperature. Then, first-quadrant M-H isotherm was recorded in step (stable-field) mode. Next, the field was decreased to zero. Since, there is no hysteresis, so one can go directly from one temperature to other. That is how M-H isotherms were recorded at various temperatures in the vicinity of  paramagnetic transition temperature, $T_{\rm{C}}$.

\paragraph{\large \bf{2. Modified Arrott plots}}\

\begin{figure*}
	\centering
	\includegraphics[width=15 cm]{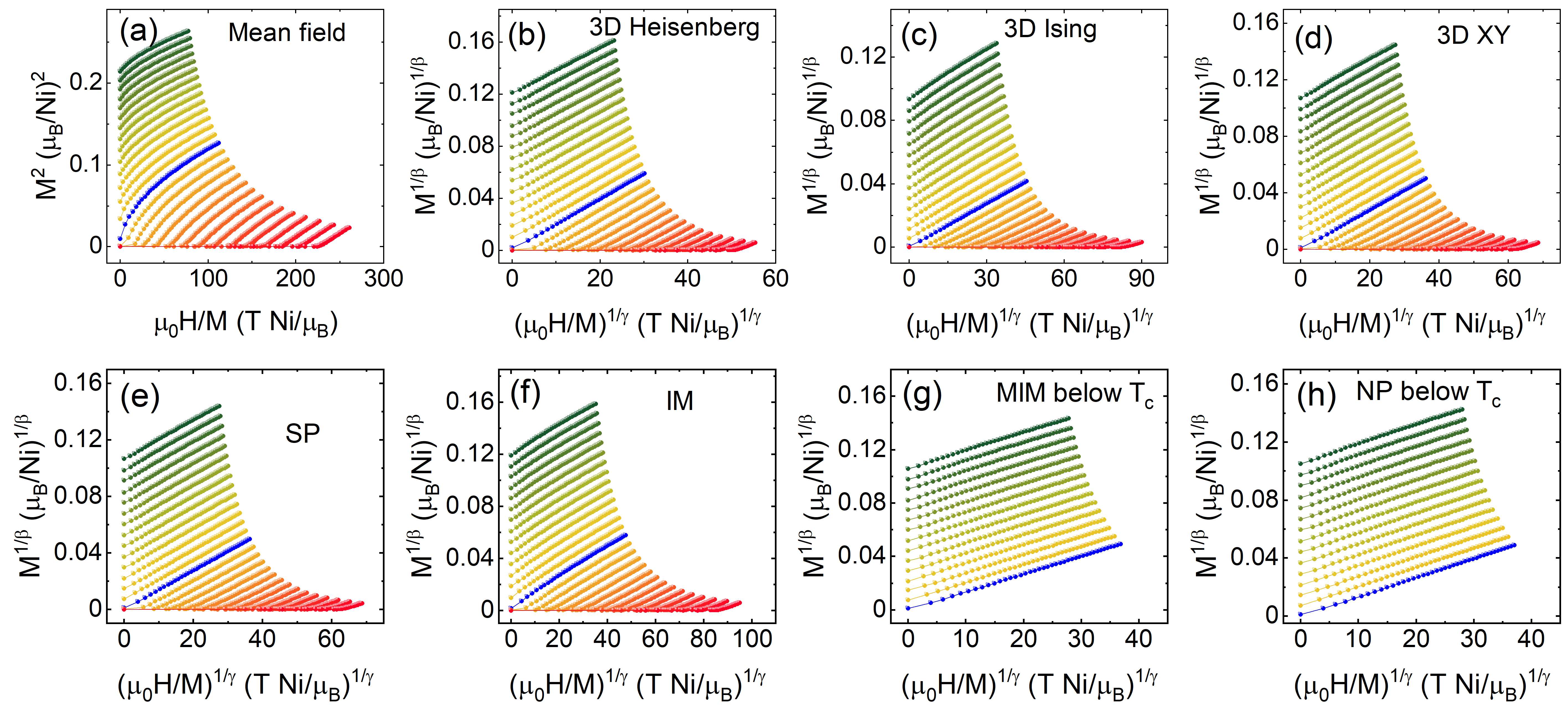}
	\caption{The modified Arrott-plots (MAPs) constructed using the $\beta$ and $\gamma$ of (a-d) standard universality classes, (e) SP, (f) IM, (g) MIM, and (h) the new proposal.}
	\label{MAPs_Ni}
\end{figure*}

\begin{figure*}
	\includegraphics[width=16 cm]{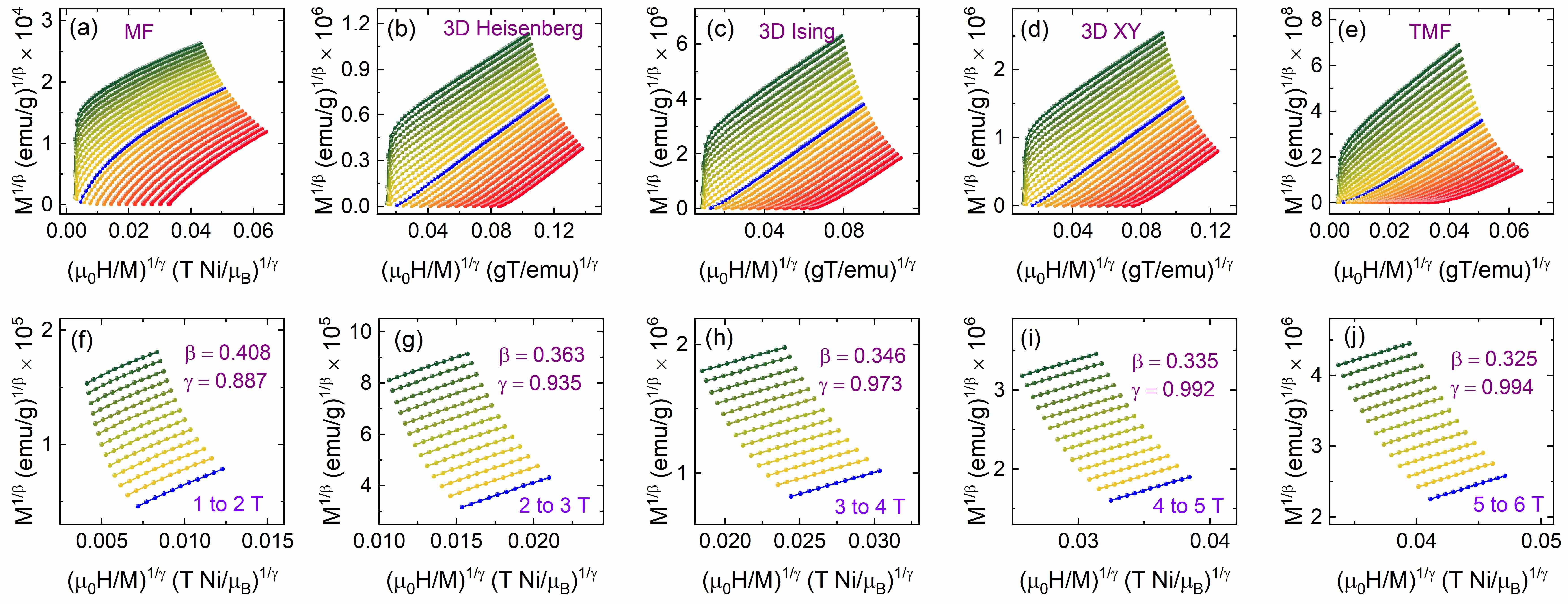}
	\caption{(a) to (e) The modified Arrott-plots (MAPs) constructed using the $\beta$ and $\gamma$ of standard universality classes for Gd. (f) to (j) The MAPs constructed using the determined $\beta$ and $\gamma$ from the proposed method for below $T_{\rm{C}}$ and in different field ranges as mentioned.}
	\label{MAPs_Gd}
\end{figure*}

In 1967, A. Arrott and J. E. Noakes \cite{PhysRevLett.19.786} developed an empirical relation between magnetization ($M$) and applied magnetic field ($\mu_0H$) using the critical exponents (CEs) $\beta$ and $\gamma$ as
\begin{equation}
(\mu_0 H/M)^{1/{\gamma}} = (T-T_{\rm{C}})/T_{\rm{C}} + (M/M_1)^{1/{\beta}},
\label{ANE}
\end{equation} 
where $M_1$ is a temperature and field-dependent constant. $T$ is the temperature at which the M-H isotherms are recorded. The plots of $(\mu_0 H/M)^{1/{\gamma}}$ vs $M^{1/{\beta}}$ is known as modified Arrott plots (MAPs), see below Figs. \ref{MAPs_Ni} and \ref{MAPs_Gd}.

\paragraph{\large \bf{3. Iteration method}}\

The concept to develop the empirical relation \ref{ANE} was to yield quasi linear behavior of MAPs [Figs. \ref{MAPs_Ni} and \ref{MAPs_Gd}] in the saturation magnetization region using a particular set of $\beta$ and $\gamma$ in the vicinity of $T_{\rm{C}}$. Later, an IM was employed to determine appropriate values of the CEs. To get appropriate $\beta$ and $\gamma$, an iteration process was employed as follows:

For a given values of $\beta$ and $\gamma$, the MAPs show quasi linear variation in the vicinity of $T_{\rm{C}}$. A linear fit to a MAP, in the saturation magnetization region, yields $slope$ and $intercept$. Thus, on comparison with Eq. \ref{ANE}, one can estimate saturation magnetization, $M_S(T)$ as
\begin{equation}
M_S(T) = (intercept)^{\beta} \,\, \text{for}\,\, T<T_{\rm{C}}\,,
\label{Ms-1}
\end{equation} 
and inverse of susceptibility, $\chi_{0}^{-1}(T)$ as
\begin{equation}
\chi_{0}^{-1}(T) = \displaystyle \left(\frac{-intercept}{slope}\right)^{\gamma} \,\, \text{for}\,\, T>T_{\rm{C}}.
\label{Sus-1}
\end{equation} 
The estimated $M_S(T)$ and $\chi_{0}^{-1}(T)$ can be further fitted with 
\begin{equation}
M_S(T) = M_0|-\epsilon|^{\beta} \,\, \text{for}\,\, \epsilon<0\,
\label{Ms}
\end{equation} 
and
\begin{equation}
\chi_{0}^{-1}(T) = h_0/M_0|\epsilon|^{\gamma} \,\, \text{for}\,\, \epsilon>0,
\label{Sus}
\end{equation} 
respectively. Here, $\epsilon=\displaystyle \frac{T-T_{\rm{C}}}{T_{\rm{C}}}$ is the reduced temperature, $M_0$ and $h_0$ are constants. The above fittings (Eqs. \ref{Ms} and \ref{Sus}) yield new values of $\beta$ and $\gamma$. Again, a new MAPs are constructed using the obtained values of $\beta$ and $\gamma$. Further, following Eqs. \ref{Ms-1} to \ref{Sus}, another set of $\beta$ and $\gamma$ are generated. The above processes will be repeated till the converged values of $\beta$ and $\gamma$ are achieved. 

\paragraph{\large \bf{4. Modified iteration method}}\

MIM yields two sets of $\beta$ and $\gamma$ on either side of $T_{\rm{C}}$ \cite{PhysRevLett.128.015703}. The process to estimate appropriate $\beta$ and $\gamma$ using MIM is as follows.\\
\begin{enumerate}
	\item First, including the M-H isotherm taken at $T_{\rm{C}}$, make two sets of M-H isotherms: one for $T\leq T_{\rm{C}}$ and another for $T\geq T_{\rm{C}}$.
	\item Take one set of M-H isotherms. Now, using $\beta$ and $\gamma$ of a standard UC, estimate normalized slopes (NSs) as discussed in the main text. One should keep in mind$-$vary $\beta$ and $\gamma$ in such a way that the intercept, estimated from the linear fit to the MAP at $T_{\rm{C}}$, should pass through origin.
	\item Take another values of $\beta$ and $\gamma$ (either by increasing or decreasing) in such a way that constructed NSs get closer to 1.
	\item Stop the process when NSs become closest to 1 within error limit. This is how one set of $\beta$ and $\gamma$ will be estimated.
	\item Repeat steps (2) to (4) for the remaining set of M-H isotherms.
\end{enumerate}
Thus, two separate sets of $\beta$ and $\gamma$ are estimated. In this article, we claim that the critical behavior remains the same on either side of $T_\mathrm{C}$. The discrepancy reported in \cite{PhysRevLett.128.015703}, may have come due to (i) closest values of NS, within error limit, above $T_\mathrm{C}$, and (ii) selection of little lower temperature range ($\approx$ 6 K) for above $T_\mathrm{C}$ as compared to the selected temperature range of 70 K, here, for Ni during computational investigation.

\begin{figure}
	\includegraphics[width=\linewidth]{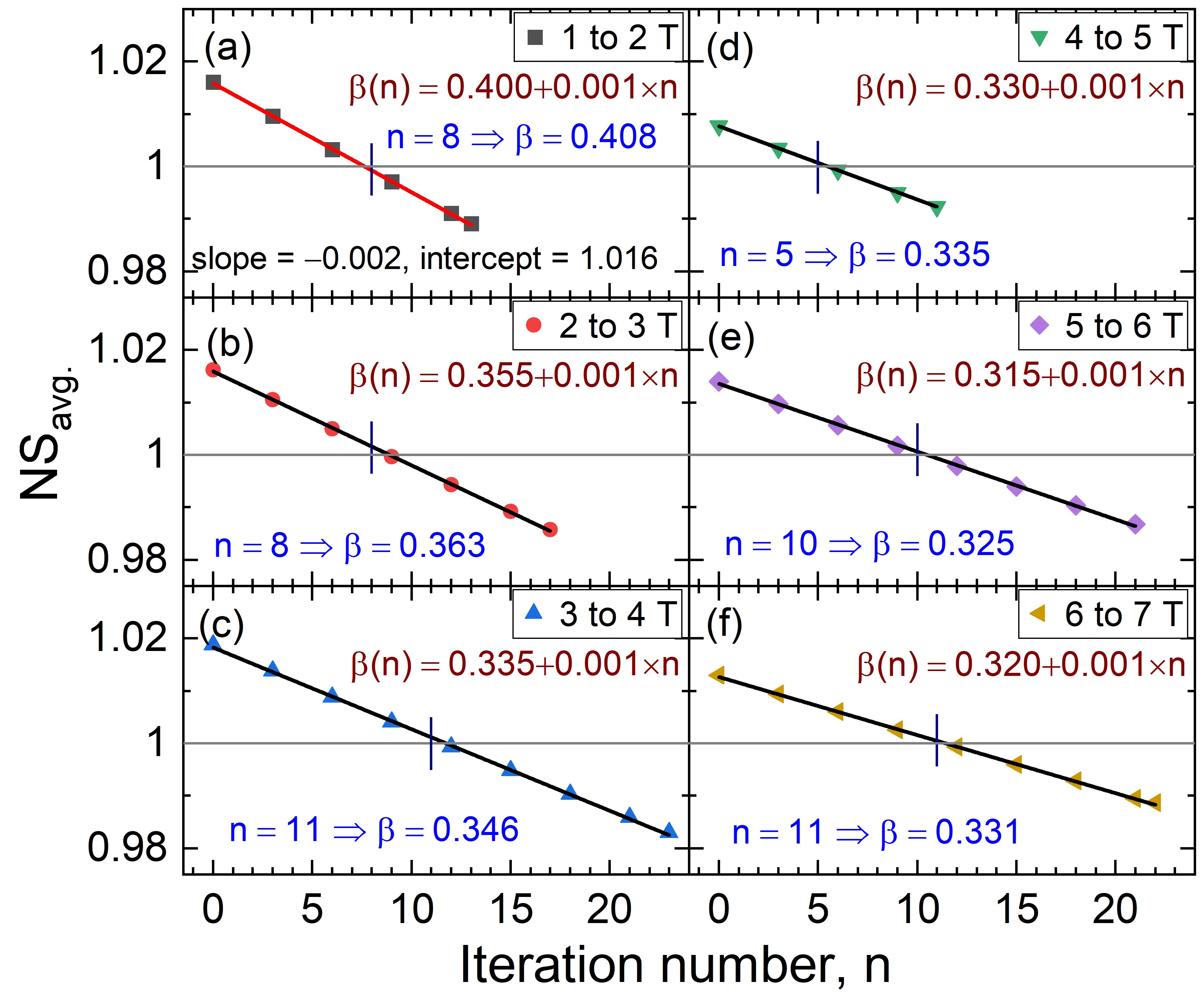}
	\caption{The variation of estimated $NS_{\rm{avg}}$, which is defined as $NS_{\rm{avg}}=\displaystyle\frac{\text{Sum of NSs}}{\text{Total number of M-H isotherms}}$, with iteration number (n), which is related with $\beta$ as $\beta(n)=\beta_i+0.001\times n$, where $\beta_i$ is different for different field ranges as mentioned in respective figures as follows: (a) field-range from 1 to 2 T and $\beta_i=0.400$, (b) field-range from 2 to 3 T and $\beta_i=0.355$, (c) field-range from 3 to 4 T and $\beta_i=0.335$, (d) field-range from 4 to 5 T and $\beta_i=0.330$, (e) field-range from 5 to 6 T and $\beta_i=0.315$, and (f) field-range from 6 to 7 T and $\beta_i=0.320$. We get final $\beta$, when $NS_{\rm{avg}}$ is approximately equal to 1 at a particular value of $n$ as mentioned in the respective figures.}
	\label{NP}
\end{figure}

\paragraph{\large \bf{5. New Proposals}}\

\textbf{a. The successive process}

In the main text, we propose a new method to get appropriate values of CEs for magnetic systems. The proposal correlates $\beta$ and $\gamma$ with $\delta$ as follows
\begin{equation}
\gamma=\beta\times(\delta-1),
\label{PR}
\end{equation}
where $\delta$ is estimated from the M-H at $T_{\rm{C}}$ using the relation
\begin{equation}
M=M_0\left(\mu_0H\right)^{\frac{1}{\delta}}, \, at \, T=T_{\rm{C}},
\label{delta}
\end{equation} 
where $M_0$ is a proportionality constant. It is necessary to estimate $\delta$ for different field range if one wants to perform field-dependent investigation of critical behavior. Our recommendation is to check if the value of $\delta$ changes with field or not. Then, depending on the situation, one need to perform critical analysis. Then, determine corresponding $\gamma$ by putting $\beta$ and $\delta$, which is determined using relation \ref{delta}, in Eq. \ref{PR}. Thus, one can get a series of $\beta$ and corresponding $\gamma$. Here, we have varied $\beta$ with iteration number, n as $\beta(n)=\beta_i+0.001\times n$. In this way, one gets error in the value of $\beta$ in third decimal place. The multiplication factor, $0.001$ is dependent on the choice of the user. Thus, we control the error in $\beta$ manually as has been discussed in the main text. It is obvious that one may get error in the magnetization during experiment. For the same, we put error in $\beta$ up to $0.003$. The reverse calculation using relation \ref{PR} is also true, \textit{i.e.}, vary $\gamma$ and determine corresponding $\beta$ as $\beta=\displaystyle \frac{\gamma}{\delta-1}$. In this way, one can control the error in $\gamma$ manually and the error in $\beta$ will depend on the imposed error in $\gamma$ and the error appearing in $\delta$.

Finally, we construct the MAPs using the sets of the CE values as discussed above and by selecting the temperature range, $T\leq T_{\rm{C}}$. The linear fit to MAPs in the selected field-range yields slopes and intercepts. As observed and has been discussed in the main text, intercepts make crossover at $T_{\rm{C}}$, which means, linear fit to MAP at $T_{\rm{C}}$ will pass through origin. After estimating $NS$, which is defined as $NS=\displaystyle\frac{\text{slope of the linear fit to MAP at }T}{\text{slope of the linear fit to MAP at }T_{\rm{C}}}$, we determine average of NS ($NS_{\rm{avg}}$) as $NS_{\rm{avg}}=\frac{\sum_{i=1}^{N} (NS)_i}{N}$, where $i$ denotes the $M-H$ isotherm at a given $T$, and $N$ is the total number of the isotherms appearing in the calculation. As shown in Fig. \ref{NP}, those values of $\beta$ and corresponding $\gamma$ (Eq. \ref{PR}) are appropriate for which $NS_{\rm{avg}}\sim1$. The details are mentioned in the main text. One can follow the same process for $T\geq T_{\rm{C}}$. However, we have shown in the main text that analysis for $T\geq T_{\rm{C}}$ is not required.

\begin{figure}
	\includegraphics[width=\linewidth]{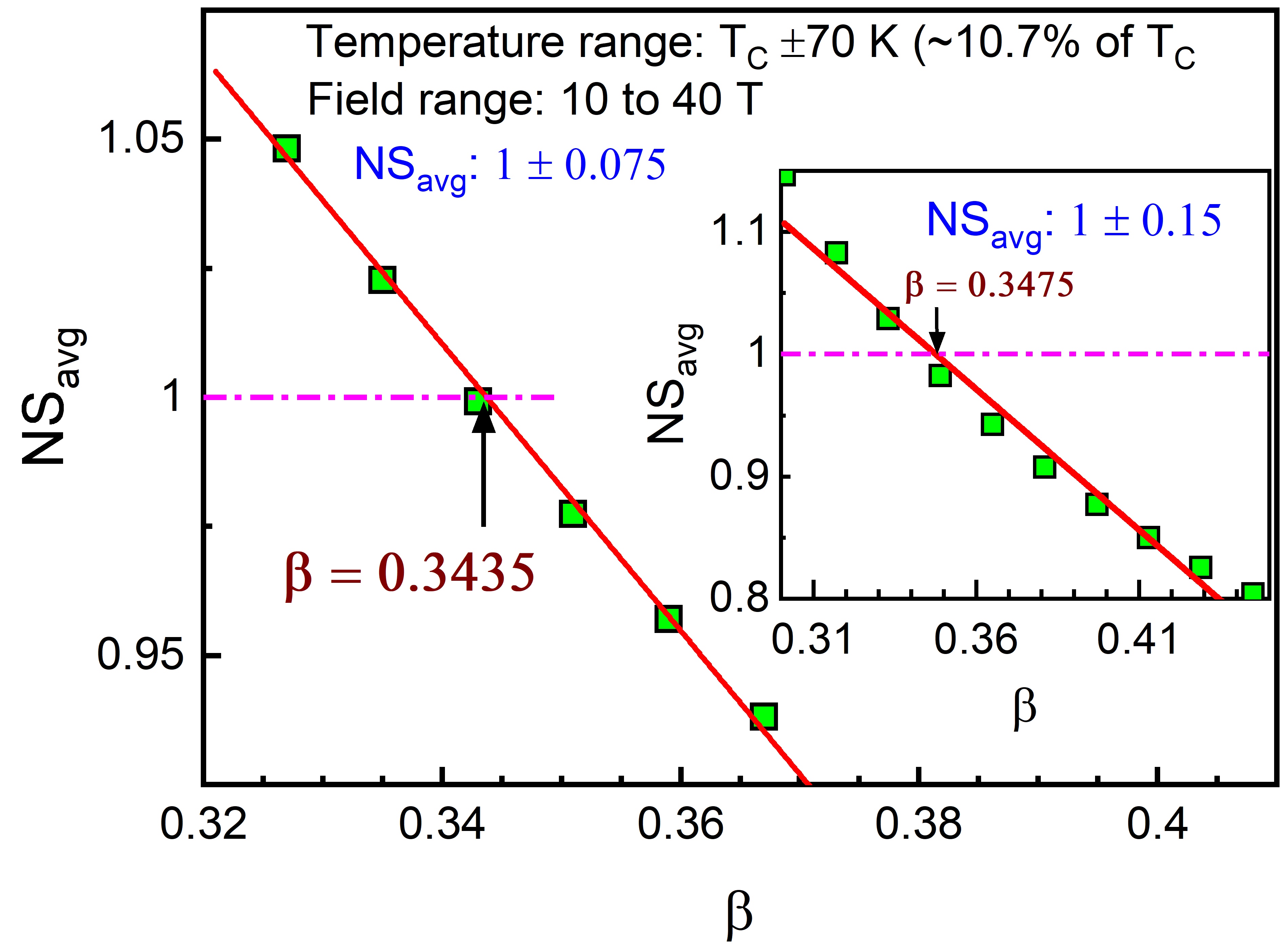}
	\caption{The variation of estimated $NS_{\rm{avg}}$ with $\beta$. The selected temperature range is $T_{\rm{C}}\pm70$ K and the field range is from 10 to 40 T. The $NS_{\rm{avg}}$ vs. $\beta$ plot show almost linear variation when values of $NS_{\rm{avg}}$ fall with $1.00\pm0.05$. The inset shows that $NS_{\rm{avg}}$ starts showing curvature when larger range of $NS_{\rm{avg}}$ is taken for analysis.}
	\label{NP_2}
\end{figure}

\begin{figure*}
	\centering
	\includegraphics[width=13cm]{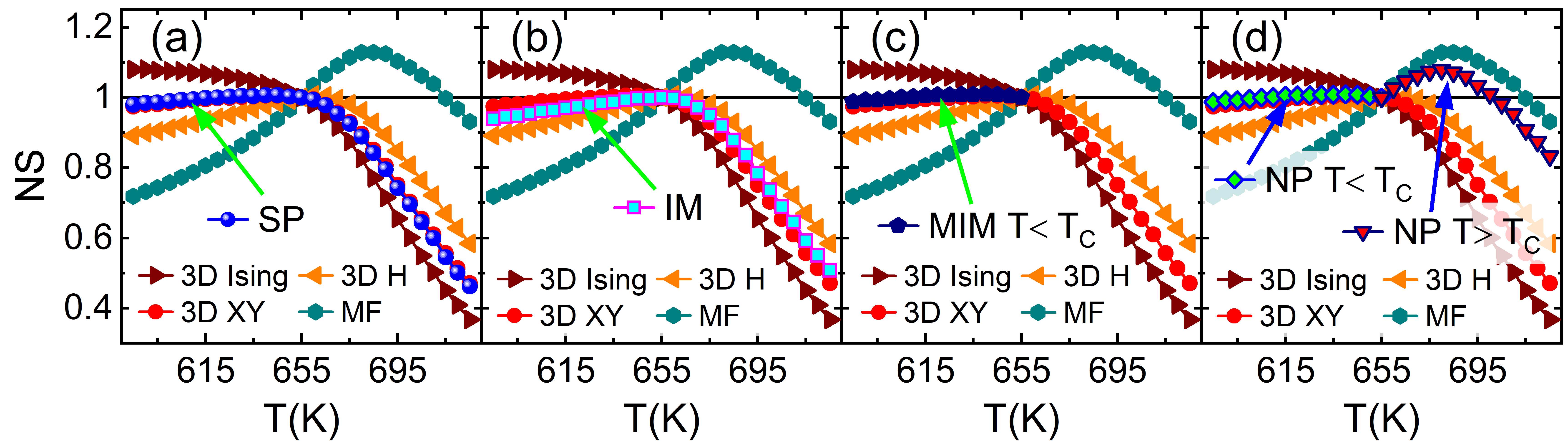}
	\caption{Normalized slope (NS) vs. temperature plots determined from the slope of the linear fit the modified Arrott-plots constructed using the critical exponents of standard UCs and the CEs determined using (a) standard procedure (SP), (b) iteration method (IM), (c) modified iteration method (MIM), and (d) our new proposal(s) (NP). The NS$\sim1$, when estimated using the CEs determined using SP, MIM, and NP for $T\leq T_{\rm{C}}$. The deviation of $NS$ from 1 started above $T_{\rm{C}}$. Also, NS deviates from 1 on either side of $T_{\rm{C}}$ when estimated using the CEs determined from IM.}
	\label{NSs_Ni}
\end{figure*}

\textbf{b. The unconventional but simpler process} 

As shown in Fig. \ref{NP}(a), the linear fit to $NS_{\rm{avg}}$ vs. $n$, yields slope$=-0.002$ and intercept$=1.016$. The linearly fitted equation can be written as
\begin{equation}
NS_{avg.}=\text{slope}\times n+\text{intercept}.
\label{Linear_Fit}
\end{equation}
Moreover, we have related $\beta$ with $n$ as $\beta(n)=\beta_i+0.001\times n$, where $\beta_i=0.400$ [Fig. \ref{NP}(a)] for the selected field range. One can get appropriate $n$ by putting $NS_{\rm{avg}}=1$, slope$=-0.002$ and intercept$=1.016$ in Eq. \ref{Linear_Fit}. Thus, from the linear fit (Eq. \ref{Linear_Fit}), we get $n=8$ using which we determine $\beta=0.408$ which matches exactly as obtained using the successive process.

In the unconventional approach, the succession process is not required, instead we select a few set of $\beta$ and corresponding $\gamma$ (correlated with Eq. \ref{PR}) in such a way that the estimated $NS_{\rm{avg}}$ falls between 0.95 and 1.05. One can go for the selection of the $NS_{\rm{avg}}$ range from 0.9 to 1.1 or more, but it should be noted that the $NS_{\rm{avg}}$ vs. $\beta$ plot should show quasi-linear variation in the vicinity of $NS_{\rm{avg}}=1$ as shown in Fig. \ref{NP_2}. Now, estimate $NS_{\rm{avg}}$ for the selected set of $\beta$ and corresponding $\gamma$, and do $NS_{\rm{avg}}$ vs. $\beta$ plot. Here, one can take a minimum sets of $\beta$ and corresponding $\gamma$ to perform a linear fit. Our, recommendation is to select at least three set of $\beta$ and corresponding $\gamma$ in such a way the estimated $NS_{\rm{avg}}$ falls on both side of $NS_{\rm{avg}}=1$. Finally, the linear fit to the plot will yield $slope$ and $intercept$. Now, using the relations, 
\begin{equation}
\beta = \frac{1-\text{intercept}}{\text{slope}}
\label{Unconventional_relation}
\end{equation}
and
\begin{equation}
\gamma = \frac{1-\text{intercept}}{\text{slope}}\times(\delta-1),
\label{Unconventional_relation2}
\end{equation}
the appropriate respective values of $\beta$ and $\gamma$ can be determined easily. We have tested this concept for the field range 10 to 40 T as shown in Fig. \ref{NP_2}. As shown, from the intersection point of the linear fit to data and the horizontal line drawn passing through $NS_{\rm{avg}}=1$, we directly get $\beta=0.3435$. By inserting the determined $slope$ and $intercept$ (Fig. \ref{NP_2}) in relation \ref{Unconventional_relation}, we get $\beta\approx0.343$, which matches exactly as determined using our proposed successive process for the estimation of CEs and the SP. However, when the range of $NS_{\rm{avg}}$ is increased up to $1.00\pm0.15$ [Inset of Fig. \ref{NP_2}], the determined value of $\beta\approx0.347$ starts deviating. And $NS_{\rm{avg}}$ vs. $\beta$ started showing curvature.

\paragraph{\large \bf{6. Normalized slope plots for Ni}}\

Figure \ref{NSs_Ni} shows the normalized slope (NS) vs. temperature plots obtained using the slope of the MAPs constructed using the CE values of the standard UCs and the values determined using standard procedure (SP). As one can see, Fig. \ref{NSs_Ni} (b), the NSs show significant deviation from $1$ on either side of $T_\mathrm{C}$ when obtained using the values of CEs as determined using the IM.

\paragraph{\large \bf{7. Field-dependent normalized slope and intercept plots for SC Gd}}\

\begin{figure*}
	\centering
	\includegraphics[width=14.5cm]{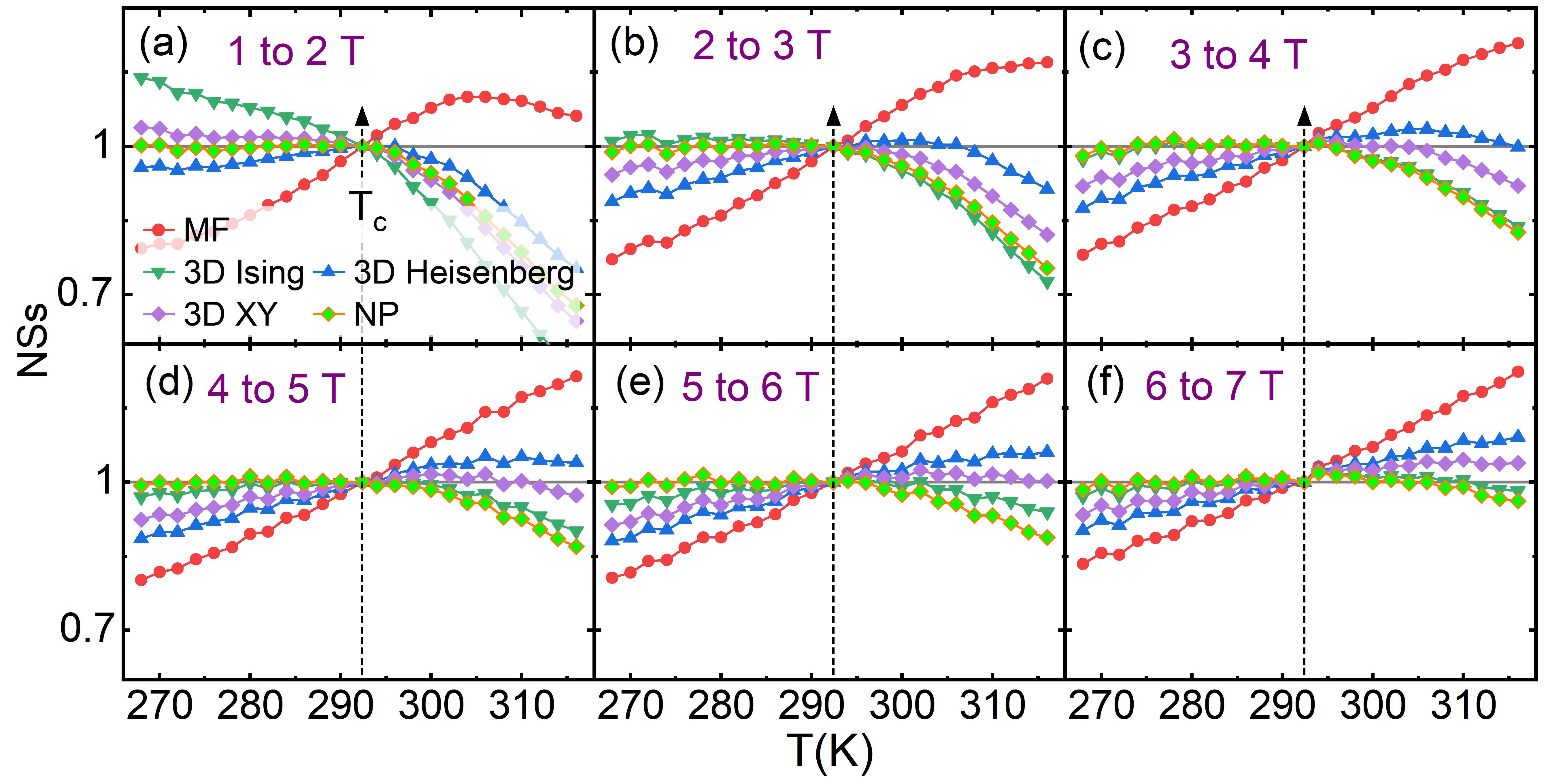}
	\caption{Normalized slopes (NSs) vs. temperature plots determined from the slope of the linear fit the modified Arrott-plots constructed using the critical exponents of standard UCs and the CEs determined using the new proposal for the field range (a) 1 to 2 T, (b) 2 to 3 T, (c) 3 to 4 T, (d) 4 to 5 T, (e) 5 to 6 T and (f) 6 to 7 T.}
	\label{NSs_Gd}
\end{figure*}

Figure \ref{NSs_Gd} represents the variation of NSs with temperature (defined above and in the main text) estimated using the CEs of standard UCs and the CEs determined using NP. As shown in Fig. \ref{NSs_Gd}, NSs make the first indication that the critical behavior of SC should belong to either 3D XY UC or 3D Heisenberg UC in low-field range. But, the determined CEs differ the guess made by NS plots. The primary guess made by the NS plots for higher field range [Figs. \ref{NSs_Gd}(b) to \ref{NSs_Gd}(f)], which is 3D Ising UC, do not depict the full physics of the system. The NS plots, constructed using the CEs determined using NP, are always closest to 1, which is the indication of the appropriate CEs. The determined CEs [Fig. \ref{NP}] do not belong to any single UC. This is the indication that one cannot depict the emerging physics directly from the NS plots using the CEs of standard UCs. Therefore, one need to perform the critical analysis to reveal the emerging physics.

\begin{figure*}
	\centering
	\includegraphics[width=14.5cm]{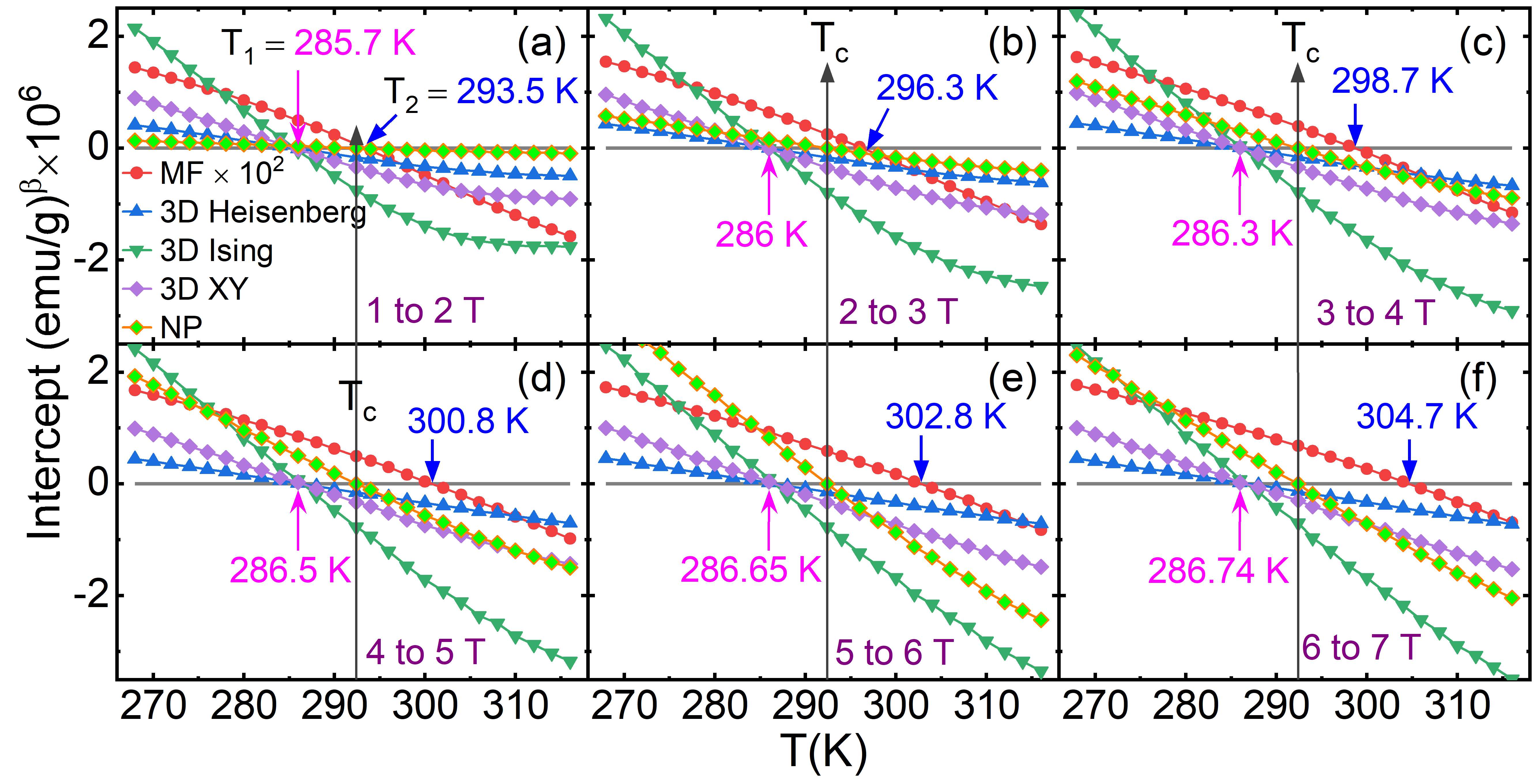}
	\caption{Intercepts vs. temperature plots determined from the intercept of the linear fit the modified Arrott-plots constructed using the critical exponents of standard UCs and the CEs determined using the new proposal for the field range (a) 1 to 2 T, (b) 2 to 3 T, (c) 3 to 4 T, (d) 4 to 5 T, (e) 5 to 6 T and (f) 6 to 7 T.}
	\label{Intercepts_Gd}
\end{figure*}

Figure \ref{Intercepts_Gd} represents the variation of Intercepts with temperature (defined above and in the main text) estimated using the CEs of standard UCs and the CEs determined using NP. Let $T_{\rm{1}}$ is the temperature where the intercepts for short-range UCs (3D Heisenberg, 3D Ising and 3D XY) make crossover. The variation of $T_{\rm{1}}$ with applied field range is almost constant. Let $T_{\rm{2}}$ is the temperature where the intercepts for long-range MF UC make crossover. The values of $T_{\rm{2}}$ show significant variation with applied field range. We have corroborated in the main text that the variation of $T_{\rm{2}}$ indicate the dominance of long-range coupling in low-field range which eventually gets weaker with applied field resulting in the coexistence of both long- and short-range coupling in the high-field range. The intercepts, estimated using the CEs determined using NP, make crossover at $T_{\rm{C}}$.

%\newpage
\paragraph{\large \bf{8. Modified Arrott plots: theoretical and experimental results}}\

Figure \ref{MAPs_Ni} represents the MAPs constructed for Ni using the CEs of standard UCs [Figs. \ref{MAPs_Ni}(a) to \ref{MAPs_Ni}(d)] and the CEs determined by standard process (SP) [Fig. \ref{MAPs_Ni}(e)], iteration method (IM) [Fig. \ref{MAPs_Ni}(f)], MIM [Fig. \ref{MAPs_Ni}(g)], and NP [Fig. \ref{MAPs_Ni}(h)] for the critical analysis. As one can see the MAPs constructed using the CEs determined from IM show slightly non-co-linear variation which is also evidenced from the NS plots [Fig. \ref{NSs_Ni}(b)] which show deviation from 1. While the MAPs constructed using the CEs determined from MIM and NP show quasi linear variation for all the isotherms for below $T_{\rm{C}}$. This is also evidenced from the NS plots [Figs. \ref{NSs_Ni}(c) and \ref{NSs_Ni}(d)] which is close to 1 for below $T_{\rm{C}}$. As observed, the linear variation of MAPs, constructed using the CEs determined from MIM and NP, for above $T_{\rm{C}}$ indicate the field-range of the linear variation decreases with increasing temperature. This investigation yields two points: (i) one should do the critical analysis in low-temperature and high-field ranges for above $T_{\rm{C}}$, and (ii) to perform critical analysis in the whole temperature range, the field-range should be varied to acquire the linear variation in the MAPs.

Figure \ref{MAPs_Gd} represents the MAPs constructed for SC Gd using the CEs of standard UCs [Figs. \ref{MAPs_Gd}(a) to \ref{MAPs_Gd}(e)] and the CEs determined by the NP for different field ranges [Figs. \ref{MAPs_Gd}(f) to \ref{MAPs_Gd}(j)] for below $T_{\rm{C}}$. The MAPs, constructed using the CEs of standard UCs, indicate that the critical behavior of SC Gd should belong to one of the short-range standard UCs. The MAPs, constructed using the CEs determined by NP, for the field range 1 to 2 T show significant but smaller deviation from quasi-linear variation. The MAPs for higher field-ranges show almost quasi-linear variation. This is the indication that the results obtained for the field range 1 to 2 T may be ambiguous. The critical analysis for above $T_{\rm{C}}$ have been ignored due to the reasons discussed in the main text while performing theoretical investigation on Ni system.

\paragraph{\large \bf{9. Scaling results}}\

\begin{figure*}
	\includegraphics[width=16 cm]{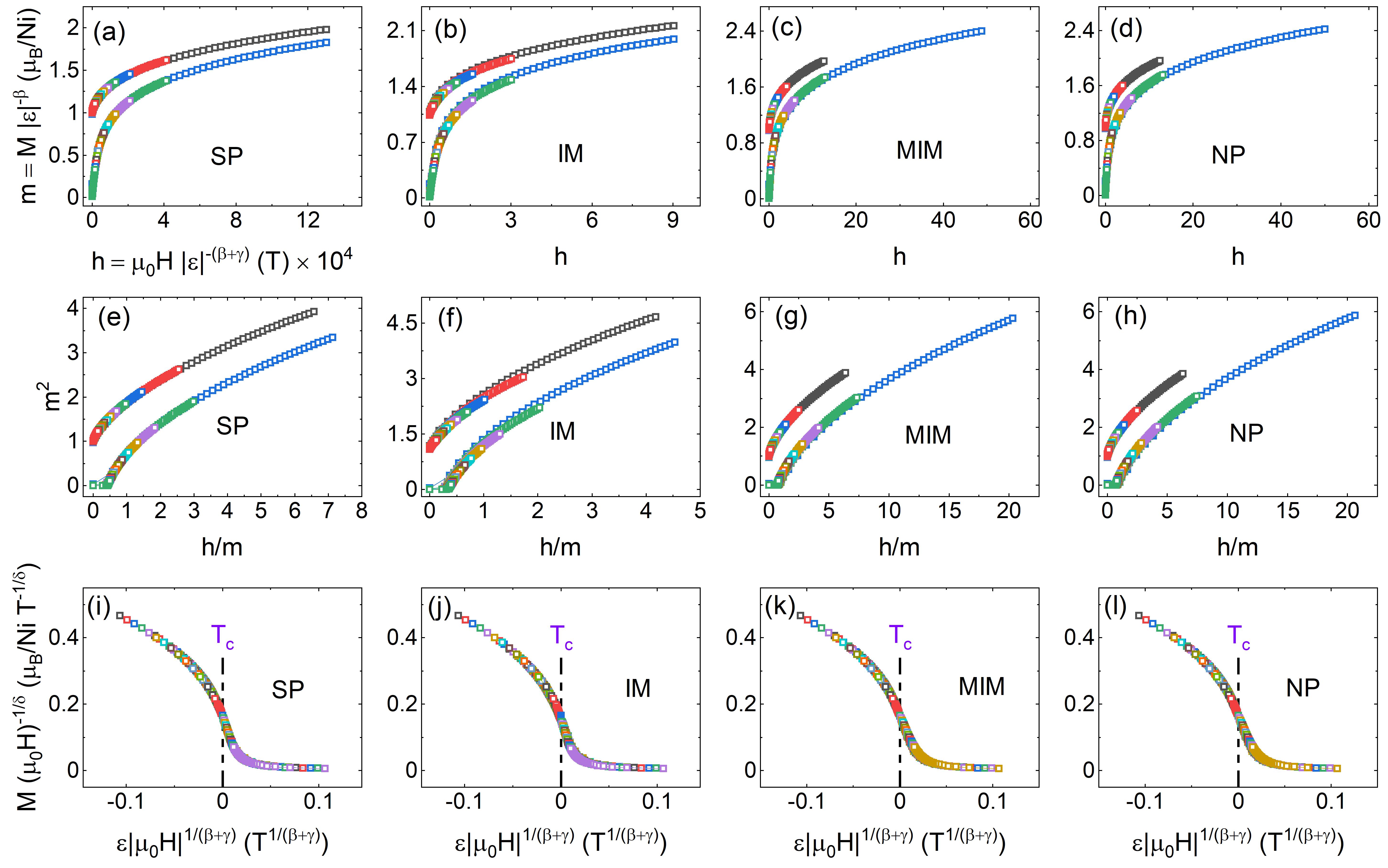}
	\caption{(a) to (d) Scaling of the M-H isotherms using the relation \ref{scaling2} and the CEs determined using standard process (SP), iteration method (IM), modified IM (MIM) and our new proposal (NP) as mentioned in the respective figures. The scaling yields renormalized magnetization, $m=M|\epsilon|^{-\beta}$ and renormalized field $h=\mu_0H|\epsilon|^{-(\beta+\gamma)}$. (e) to (h) The renormalized Arrott-plots (RAPs), $m^2$ vs. $h/m$, for the CEs determined from SP, IM, MIM and NP as mentioned in the respective figures. (i) to (l) Scaling of the M-H isotherms using the relation \ref{scaling1} and the CEs determined using SP, IM, MIM and NP as mentioned in the respective figures.}
	\label{Scaling_Ni}
\end{figure*}

The scaling equation of state, for magnetic systems, takes the form \cite{RevModPhys.71.S358, KAUL19855}
\begin{equation}
\frac{\mu_0H}{M^{1/\delta}} = \tilde{h}\left(\frac{\epsilon}{M^{1/\beta}}\right),
\label{scaling1}
\end{equation}
where $\tilde{h}(x)$ is the scaling function. Thus, the relation \ref{scaling1} states that for appropriate CEs and $T_{\rm{C}}$, the plots of $M(\mu_0H)^{1/\delta}$ vs. $\epsilon$$(\mu_0$H)$^{-\frac{1}{\beta+\gamma}}$ should correspond to a universal curve onto which all the isotherms collapse. The Eq. \ref{scaling1} may be formally converted as
\begin{equation}
M(\mu_0H,\epsilon)|\epsilon|^{-\beta}= f_{\pm}\left(\frac{\mu_0H}{\epsilon^{\beta+\gamma}}\right),
\label{scaling2}
\end{equation}
where $f_+$ and $f_-$ are defined for $T>T_{\rm{C}}$ and $T<T_{\rm{C}}$, respectively. The relation \ref{scaling2} states $-$ for appropriate values of $\beta$ and $\gamma$, the plots of renormalized magnetization, $m=M(H,\epsilon)|\epsilon|^{-\beta}$ vs. renormalized field, $h=\mu_0H|\epsilon|^{-(\beta+\gamma)}$ should correspond to collapse of the isotherms on two separate universal curves for below and above $T_{\rm{C}}$. Equation \ref{scaling2} is most commonly used relation to verify the appropriateness of the obtained CEs. It has been proven \cite{PhysRevLett.128.015703} that these scaling relations (Eqs. \ref{scaling1} and \ref{scaling2}) can absorb significantly large errors appearing in the determined values of CEs. So, a new and appropriate relation is needed to check the reliability of determined CEs. However, Widom relation, $\delta= 1+\displaystyle \frac{\gamma}{\beta}$, may be used as a primary tool to check the reliability and accuracy of determined CEs.

\begin{figure*}
	\includegraphics[width=16 cm]{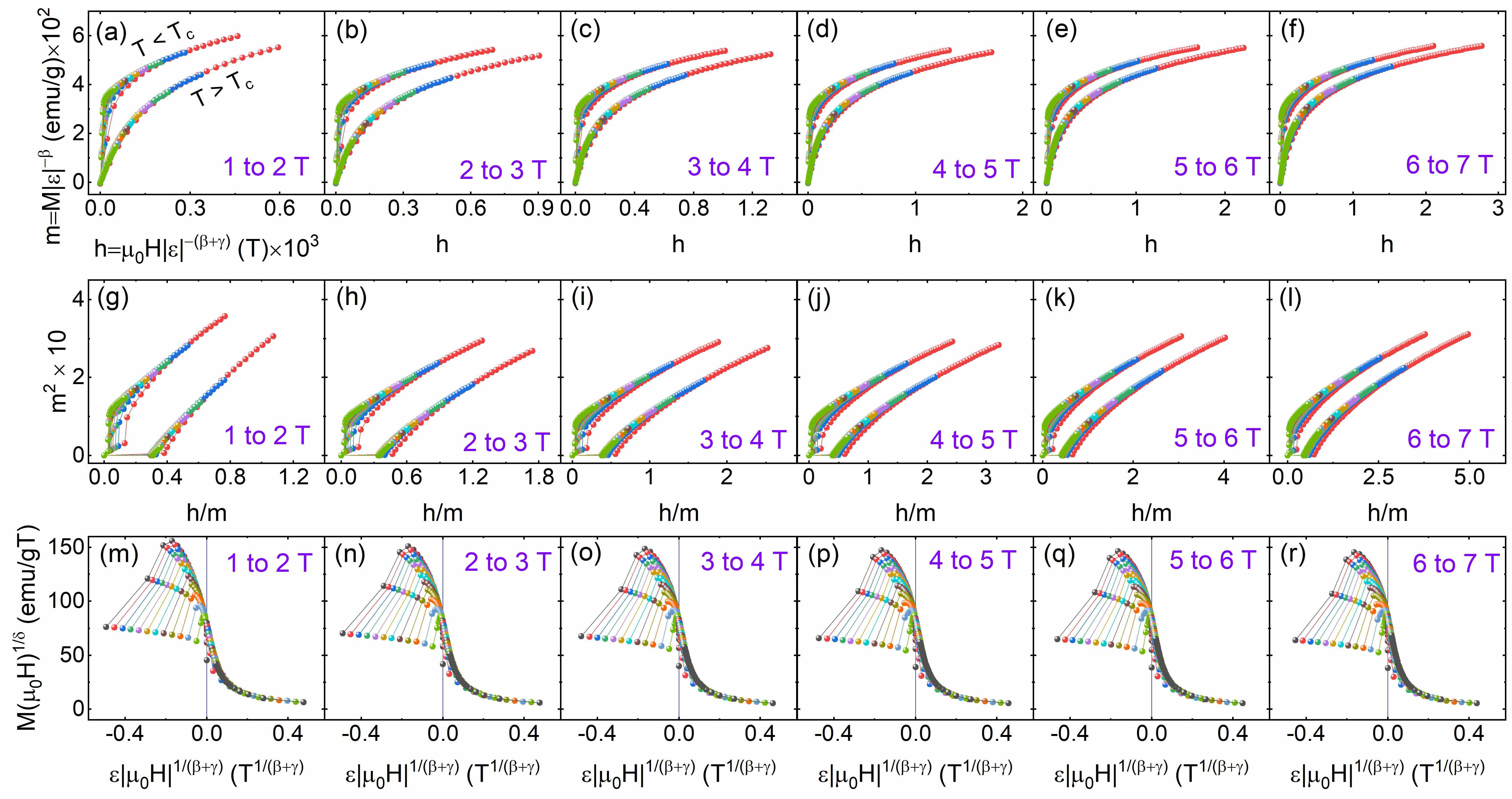}
	\caption{(a) to (f) Scaling of the M-H isotherms using the relation \ref{scaling2} and the CEs determined using our new proposal (NP) for different field-ranges as mentioned in the respective figures. The scaling yields renormalized magnetization, $m=M|\epsilon|^{-\beta}$ and renormalized field, $h=\mu_0H|\epsilon|^{-(\beta+\gamma)}$. (g) to (l) The renormalized Arrott-plots (RAPs), $m^2$ vs. $h/m$, for the CEs determined from NP for the different field-ranges as mentioned in the respective figures. (m) to (r) Scaling of the M-H isotherms using the relation \ref{scaling1} and the CEs determined using NP for the different field-ranges as mentioned in the respective figures.}
	\label{Scaling_Gd}
\end{figure*}

Figures \ref{Scaling_Ni}(a) to \ref{Scaling_Ni}(d) show the plots of $m$ vs. $h$ constructed using the CEs determined by SP, IM, MIM and NP for Ni system. As one can see clearly that the $m$ vs. $h$ plots using the CEs determined by IM [Fig. \ref{Scaling_Ni}(b)] show maximal deviation from overlapping on two separate isotherms. The $m$ vs. $h$ using the CEs determined by SP [Fig. \ref{Scaling_Ni}(a)] have perfect collapse on two universal curves for the whole range of the M-H isotherms. Similar perfect collapse on two universal curves for the whole range of the M-H isotherms have been observed using the CEs determined by MIM [Fig. \ref{Scaling_Ni}(c)] and NP [Fig. \ref{Scaling_Ni}(d)] for below $T_{\rm{C}}$. We have used the same CEs, determined by MIM and NP for below $T_{\rm{C}}$, to construct the $m$ vs. $h$ plots for above $T_{\rm{C}}$, which show slight deviation from collapsing on a universal curve. Figures \ref{Scaling_Ni}(e) to \ref{Scaling_Ni}(h) are the renormalized Arrott-plots (RAPs) constructed using the CEs determined by SP, IM, MIM and NP. The perfect overlapping is closely observable for the CEs determined by SP, and MIM and NP for below $T_{\rm{C}}$. The RAPs constructed using the CEs determined by IM, MIM for above $T_{\rm{C}}$ and NP for above $T_{\rm{C}}$ show slightly large deviation from collapsing on universal curve. These are the indication that the CEs determined using SP, and MIM and NP for below $T_{\rm{C}}$ are appropriate to reveal the properties of a system under investigation. In fact, the CEs determined from the above three mentioned methods (SP, MIM and NP) match significantly to each other within error limit. However, no concrete conclusion can be made from the use of relation \ref{scaling1} because the constructed plots using the CEs determined by all methods overlap significantly onto a single universal curves as shown in Figs. \ref{Scaling_Ni}(i) to \ref{Scaling_Ni}(l). From these investigations, one important information is coming out $-$ the overlapping of all the isotherms on universal curves for entire field-range is the indication of the existence of single exchange coupling in the system, which was intentionally incorporated during simulation.

Figure \ref{Scaling_Gd} represents the scaling plots and RAPs for SC Gd. One should note that the critical analysis on SC Gd have been carried out only for below $T_{\rm{C}}$ but we have plotted the above $T_{\rm{C}}$ isotherms also using the same CEs as determined for below $T_{\rm{C}}$. The plots shown in Fig. \ref{Scaling_Gd}, have been constructed using the CEs determined by NP. The $m$ vs. $h$ plots show nice overlapping in the field-ranges which were used to determine CEs for below $T_{\rm{C}}$. However, in the low field range (approximately below 0.5 T) the $m$ vs. $h$ plots [Figs. \ref{Scaling_Gd}(a) to \ref{Scaling_Gd}(f)] show deviations from collapsing on the universal curves. In the similar fashion, the RAPs [Figs. \ref{Scaling_Gd}(g) to \ref{Scaling_Gd}(l)] also sow overlapping in the high field range while deviations are observable in the low field range. The scaling plots [Figs. \ref{Scaling_Gd}(m) to \ref{Scaling_Gd}(r)], constructed using the relation \ref{scaling1}, also show significant deviations in the low-field range mainly below $T_{\rm{C}}$. The reported scaling results using Eq. \ref{scaling1} do not show deviation from collapsing on single universal curve. The reason behind this could be the skipping of the low-field data. However, from the deviation in the low-field region, we state that this is the signature of the competing energies along with symmetric exchange interaction in SC Gd.

\paragraph{\large \bf{10. Table of critical exponents of standard 3D models, and the critical exponents determined for Ni and SC Gd estimated using various methods}}\

\begin{table*}
	\centering
	\caption{The list of CEs ($\beta$, $\gamma$, $\delta$ and $\alpha$) of standard 3D universality classes and the CEs determined using various methods as discussed above and presented in the main text. $Ref.$: $References$, Tech: Techniques, RG: renormalization group, TMF: tricritical mean-field, H: Heisenberg. ME: magnetic entropy, PM: proposed method, Expt.: experiment, $d$: space dimensionality, and $n$: spin dimensionality.}
	%	\label{table1}
	%	\begin{ruledtabular}
	\begin{tabular}{|c|c|c|c|c|c|c|c|c|}
		\hline
		\multirow{2}{*}{Model/} & \multirow{2}{*}{Ref/}& \multirow{2}{*}{Method}& \multicolumn{4}{c|}{Critical Exponents} &\multicolumn{2}{c|}{Dimensionality} \\ 
		\cline{4-9}
		Sample &Field-range & & {$\beta$} & {$\gamma$} & {$\delta$} & {$\sigma$} & \hspace*{3mm} {$d$} \hspace*{3mm} & {$n$} \\
		\hline
		{Mean-field} & \cite{PhysRevB.65.214424, KAUL19855} & Theory & {0.500} & {1.000} & {3.000} & {1.500} & {3} & {$-$}\\
		\hline
		{TMF} & \cite{PhysRevB.65.214424} & Theory & {0.250} & {1.000} & {5.000} & {1.500} & {3} & {$-$}\\
		\hline
		{3D Ising} & \cite{PhysRevB.65.214424, KAUL19855} & Theory & {0.325} & {1.237} & {4.800} & {1.960} & {3} & {1}\\
		\hline
		{3D XY} & \cite{PhysRevB.65.214424, KAUL19855} & Theory & {0.345} & {1.316} & {4.800} & {1.960} & {3} & {2}\\
		\hline
		{3D H} & \cite{PhysRevB.65.214424, KAUL19855} & Theory & {0.365} & {1.386} & {4.800} & {1.960} & {3} & {3}\\
		\hline
		{Ni} & This Paper & Theory (SP) & {0.344(2)} & {1.315(8)} & {4.823(15)} & {1.920(8)} & {3} & {2}\\
		\hline
		{Ni} & This Paper & Theory (IM) & {0.362(1)} & {1.222(10)} & {4.376(37)} & {1.877(14)} & {3} & {1}\\
		\hline
		{Ni} & This Paper & Theory (MIM) & {0.343(3)} & {1.309(8)} & {4.816(57)} & {1.931(8)} & {3} & {2}\\
		\hline
		{Ni} & This Paper & Theory (NP) & {0.342(3)} & {1.307(12)} & {4.822(69)} & {1.911(13)} & {3} & {2}\\
		\hline
		{SC Gd} & 1 to 2 T & Expt. (NP) & {0.408(3)} & {0.887(14)} & {3.174(24)} & {1.247(49)} & {3} & {$-$}\\
		\hline
		{SC Gd} & 2 to 3 T & Expt. (NP) & {0.363(3)} & {0.935(10)} & {3.576(13)} & {1.377(22)} & {3} & {$-$}\\
		\hline
		{SC Gd} & 3 to 4 T & Expt. (NP) & {0.346(3)} & {0.973(9)} & {3.811(10)} & {1.453(16)} & {3} & {$-$}\\
		\hline
		{SC Gd} & 4 to 5 T & Expt. (NP) & {0.335(3)} & {0.992(9)} & {3.961(11)} & {1.487(15)} & {3} & {$-$}\\
		\hline
		{SC Gd} & 5 to 6 T & Expt. (NP) & {0.325(3)} & {0.994(8)} & {4.060(7)} & {1.490(13)} & {3} & {$-$}\\
		\hline
		{SC Gd} & 6 to 7 T & Expt. (NP) & {0.331(3)} & {1.055(10)} & {4.186(12)} & {1.583(14)} & {3} & {$-$}\\
		\hline

	\end{tabular}
	%	\end{ruledtabular}
	\label{Table_Critical exponents}
\end{table*}

\pagebreak
%\section*{References}

\end{document}